\documentclass[11pt,a4paper]{article}
\pdfoutput = 1

\usepackage{jcappub}
\usepackage[utf8]{inputenc}
\usepackage{amsmath,amssymb}
\usepackage{graphicx}
\usepackage{hyperref}
\usepackage{natbib}
\usepackage{xcolor}
\usepackage{physics}
\usepackage{derivative}
\usepackage{subcaption}

\newcommand{\edit}[1]{#1}

\newcommand{\bq}{\textbf{q}}
\newcommand{\bx}{\textbf{x}}
\newcommand{\bk}{\textbf{k}}

\newcommand{\bv}{\textbf{v}}

\newcommand{\bPsi}{\boldsymbol{\Psi}}

\newcommand{\half}{\frac{1}{2}}

\newcommand{\dDelta}{\dot{\Delta}}

\newcommand{\leg}[2]{\mathcal{P}_{#1}(#2)}
\newcommand{\aleg}[3]{\mathcal{P}_{#1}^{#2}(#3)}

\def \dtq{\int d^3 \bq \ }
\def\ie{{\em i.e.}~}

\newcommand{\avg}[1]{\ensuremath{\left\langle #1 \right\rangle}}

\newcommand{\tj}[6]{ \begin{pmatrix}
   #1 & #2 & #3 \\
   #4 & #5 & #6 
  \end{pmatrix}}

\def\kMpc{\, h \, {\rm Mpc}^{-1}}
\def\icMpc{\, h^3 \, {\rm Mpc}^{-3}}

\def\hq{\hat{q}} 
\def\hn{\hat{n}}
\def\hk{\hat{k}}

\title{Redshift-Space Distortions in Lagrangian Perturbation Theory}

\author[a]{Shi-Fan Chen}
\author[b,c,e]{Zvonimir Vlah}
\author[d,e]{Emanuele Castorina}
\author[a]{Martin White}

\affiliation[a]{Department of Physics, University of California,
Berkeley, CA, USA}
\affiliation[b]{Kavli Institute for Cosmology, University of Cambridge, Cambridge, UK.}
\affiliation[c]{Department of Applied Mathematics and Theoretical Physics, University of Cambridge, Cambridge, UK.}
\affiliation[d]{Dipartimento di Fisica `Aldo Pontremoli', Universita' degli Studi di Milano, Milan, Italy}
\affiliation[e]{Theoretical Physics Department, CERN, 1211 Geneva 23, Switzerland}

\emailAdd{shifan\_chen@berkeley.edu}
\emailAdd{zv217@cam.ac.uk}
\emailAdd{emanuele.castorina@unimi.it}
\emailAdd{mwhite@berkeley.edu}

\abstract{
We present the one-loop 2-point function of biased tracers in redshift space computed with Lagrangian perturbation theory, including a full resummation of both long-wavelength (infrared) displacements and \edit{associated} velocities.  The resulting model accurately predicts the power spectrum and correlation function of halos and mock galaxies from two different sets of N-body simulations at the percent level for quasi-linear scales, including the damping of the baryon acoustic oscillation signal due to the bulk motions of galaxies.  We compare this full resummation with other, approximate, techniques including the moment expansion and Gaussian streaming model.  We discuss infrared resummation in detail and compare our Lagrangian formulation with the Eulerian theory augmented by an infrared resummation based on splitting the input power spectrum into ``wiggle'' and ``no-wiggle'' components.  
We show that our model is able to recover unbiased cosmological parameters in mock data encompassing a volume much larger than what will be available to future galaxy surveys.
We demonstrate how to efficiently compute the resulting expressions numerically, making available a fast Python code capable of rapidly computing these statistics in both configuration and Fourier space.
}

\begin{document}

\maketitle

\section{Introduction}
\label{sec:intro}
The measured redshifts of galaxies receive a contribution proportional to the relative velocity between the observer and the emitting source. This term breaks the isotropy of space, as it singles out the observer location as special, and causes a distinct anisotropic pattern in the clustering statistics of biased tracers known as Redshift Space Distortions (RSD) \cite{Kai87,Ham92,Pea99,Dod03}.  Being a probe of the velocity field, RSD contain extra cosmological information compared to the density field only, and they have been shown to be a useful probe of modified gravity models \cite{Guzzo:2008,Wei13,Joyce15,Amendola18,Alam20}.  Current and upcoming spectroscopic redshift surveys, like DESI \cite{DESI} and Euclid \cite{Euclid}, will provide measurements of the power spectrum of galaxies with much better precision than currently available, making the modeling of RSD of paramount importance to achieve their science goals. Within the framework of Perturbation Theory (PT) several different approaches have been put forward to compute clustering statistics in redshift space. We can divide them into two main categories: Eulerian PT (EPT) methods, where density and velocity fields are the relevant degrees of freedom (dof), and Lagrangian PT (LPT) methods, where the displacements of dark matter particles and galaxies are the fundamental dof from which observables are computed\footnote{In both cases we will only consider the effective field theory (EFT) approach to PT, see \cite{BNSZ12,CHS12,PorSenZal14,VlaWhiAvi15,Sen14,Ang15} and references therein.}. This work focuses on the latter, but more generally one of our main goals is to clarify the relation between the two approaches to RSD. 

LPT has a long history, especially in the context of RSD \cite{Mat08a,CLPT,Wan14,Whi14,VlaCasWhi16,VlahWhite19,ChenVlahWhite19}, and LPT models of the galaxy correlation function, \ie in position space, have been successfully applied to data (see e.g.\  refs.~\cite{Reid12,Samushia14,Satpathy17,Tamone20,Zhang20,Bautista21} for a sampling of the literature). 
In Fourier space, while RSD in EPT can be straightforwardly implemented \cite{delaBella:2017qjy,delaBella:2018fdb,Desjacques18,Ivanov20,DAmico20,Tomlinson2020}, LPT has posed a number of technical difficulties that have only recently been overcome \cite{VlahWhite19}. 
Using an expansion in the moments of the density-weighted pairwise velocities, ref.~\cite{ChenVlahWhite19} presented a derivation of the $1$-loop redshift space galaxy power spectrum in LPT, finding good agreement in comparison to simulated data.

The main goal of this paper is to compute the $1$-loop redshift space power spectrum in LPT via direct evaluation of the integrals, without employing the moment expansion (MOME) \cite{SelMcD11,Vla12,Vla13,VlahWhite19}. This is not only a practical choice, but, for example, it will allow us to clarify the effect on the power spectrum of various resummation schemes for the long-wavelength (IR) displacement modes \cite{SenZal15,Trevisan2017,VlaWhiAvi15,Baldauf15,Vlah16,Blas16,Ding18,Ivanov18}.

In EPT, IR-resummation is performed a posteriori, after the $1$-loop power spectrum is computed, and different procedures have been discussed in the literature.  In redshift space in particular, a number of approximations have been employed to render the calculation more tractable \cite{Lewandowski2015a,Perko2016,Ding18,Ivanov18}.
In the current implementation of MOME in LPT \cite{ChenVlahWhite19}, only the long wavelength displacements are resummed, leaving the long wavelength velocity effects un-resummed.  If desired, these long wavelength velocity effects can
also be resummed in a posteriori way, as is done in EPT, and as we show \edit{in Appendix~\ref{app:kir}}. \edit{Alternatively, by truncating the configuration-space velocity cumulants at second order, the Gaussian streaming model can be used to approximately resum velocities from linear (Gaussian) modes at the expense of neglecting some one-loop contributions\footnote{See, e.g.\ Appendix B of ref.~\cite{ChenVlahWhite20}.} \cite{VlaCasWhi16}. }

The direct LPT approach allows us to efficiently resum the long displacement contributions without relying on any of the above-mentioned approximations. Some of these different choices for IR-resummation can lead to different behaviours of the power spectra at small (UV) scales. However, as we shall discuss further below, these differences can be associated to the different perturbative expansion parameters they employ. Moreover, even though these differences arise from long wavelength displacements, which are under perturbative control, these residual contributions are also approximately degenerate with, and can thus be absorbed by, the free coefficients of the effective theory.

The results presented in this paper complement the existing literature on the one-loop LPT power spectrum in $\Lambda$CDM cosmologies referenced above.
Additionally, our work provides the analytical machinery to better understand the performance of forward model or density field reconstruction algorithms in redshift space based on Lagrangian displacements \cite{Zhu2017,Modi2019,Schmittfull2020}.

This paper is structured as follows. Section \ref{sec:lpt} will introduce the notation, Section \ref{sec:pk_rsd} the relevant equations for the computation of the one-loop RSD power spectrum of biased tracers, whose evaluation is discussed in Section \ref{sec:numeric}. Numerical fits to N-body simulations and mock catalogs are presented in Section \ref{sec:results}, as well as comparisons between the different PT methods. This section also demonstrates that the model is able to recover unbiased estimates of cosmological parameters in a ``blind'' challenge.  Our conclusions are presented in Section \ref{sec:conclude}. 
A number of technical points are relegated to Appendices \ref{app:kir}, \ref{app:mi} and \ref{app:mii}.

\section{Overview of Lagrangian Perturbation Theory}
\label{sec:lpt}

Our goal in this section is to give a quick overview of Lagrangian perturbation theory (LPT) as pertains to this paper, both as a review and to establish our notation and conventions. The reader is referred to the references in the introduction for further details, and especially to refs.~\cite{Mat08a,Mat08b,CLPT,Whi14,VlaWhiAvi15,VlaCasWhi16,ChenVlahWhite20} whose notations we adopt.

Within the Lagrangian picture the gravitational evolution of large-scale structure is described via the of motion fluid elements starting at initial (Lagrangian) positions $\bq$ with trajectories given by $\bx(\bq,t) = \bq + \bPsi(\bq,t)$. The Lagrangian displacements, $\bPsi$, obey the equation of motion $\ddot{\bPsi}(\bq) + \mathcal{H} \dot{\bPsi}(\bq) = -\nabla_{\bx} \Phi(\bx)$, where dots indicate derivatives with respect to the conformal time, and the gravitational potential $\Phi$ is in turn sourced by the matter overdensity $\delta_m$ given by
\begin{align}
    1 + \delta_m(\bx) = \dtq \delta_D(\bx - \bq - \bPsi(\bq)) \quad , \quad (2\pi)^3 \delta_D(\bk) + \tilde{\delta}_m(\bk) = \dtq e^{i \bk \cdot (\bq + \bPsi)}
    \label{eqn:deltam}
\end{align}
via Poisson's equation. In LPT these quantities are solved for order-by-order in the initial conditions $\delta_0(\bq)$, such that the displacements are given by $\bPsi = \bPsi^{(1)} + \bPsi^{(2)} + \bPsi^{(3)} + ...$ Of particular interest is the linear solution $\bPsi^{(1)} = - D(z) \nabla^{-1}_{\bq} \delta_0(\bq)$, also known as the Zeldovich approximation. The specific forms of the higher-order solutions are given for example in refs.~\cite{Mat15,ZheFri14,Rampf12}. These solutions contain parametrizeable dependences on small-scale physics which are captured by including additional effective-theory counterterms \cite{PorSenZal14,VlaWhiAvi15}.

In this paper we will be primarily interested in the clustering of biased tracers of matter like galaxies which are the target of galaxy redshift surveys. Within the Lagrangian framework biased tracers are modeled as functionals of the initial conditions $F[\delta_0](\bq)$ at their Lagrangian positions $\bq$ and advected along with the matter fluid, such that their observed overdensities are given by number conservation to be
\begin{equation}
1 + \delta_g(\bx) = \dtq F(\bq)\ \delta_D(\bx - \bq - \bPsi).   
\end{equation}
The bias functional $F(\bq)$ is a local function of the initial conditions with effective corrections, and at one-loop order in the power spectrum includes linear and quadratic density bias, shear and third-order contributions as well as effective corrections like derivative bias ($\propto \nabla^2 \delta^2$) similar to the dynamical ones described at the end of the previous paragraph. We will furthermore assume that cold dark matter and baryons can be treated as a single fluid and therefore discard bias operators \edit{and dynamics} proportional to relative density and velocity of the different fluids \cite{Ang15,Beu16,Sch16,ChenCasWhi19,Barreira2019,Khoraminezhad:2020zqe,Rampf20}. These extra terms are thought to be small and \edit{therefore need only be} implemented at leading order in the (relative) displacements, as discussed in ref.~\cite{ChenCasWhi19}.  Our conventions follow Equation 5.1 in ref.~\cite{ChenVlahWhite20}.

Finally, galaxy surveys determine the line-of-sight (LOS) position of observed galaxies via redshifts $z$ whose cosmological and peculiar-velocity contributions are degenerate. These redshift-space distortions (RSD) can be accounted for within LPT by boosting displacements along the LOS direction $\hn$ by their corresponding velocities $\bPsi^s = \bPsi + (\hn \cdot \bv) \hn / \mathcal{H}$, where $\mathcal{H}$ is the conformal Hubble parameter. We will use the superscript $s$ to refer to vectors boosted into redshift space throughout this work. Within the Einstein-de Sitter approximation (EdS) we have the further simplification that
\begin{equation}
\label{eqn:PsiRSD}
    \bPsi^{s,(n)} = \bPsi^{(n)} + nf\ (\hn \cdot \bPsi^{(n)})\ \hn \equiv R^{(n)} \bPsi^{(n)},
\end{equation}
where the matrix $R^{(n)}_{ij} = \delta_{ij} + nf\ \hn_i \hn_j$ and $f$ is the linear growth rate. We will operate within the EdS approximation for the remainder of the paper, and further make the distant observer approximation such that $\hat{n}$ is the same for each galaxy.  These approximations are known to be quite good in the limit of high redshifts and on scales where higher order perturbation theory is most applicable.  A discussion of violations of these approximations within the LPT context can be found in refs.~\cite{Rampf15,FasVla16,FujitaVlah20,CasWhi18a,CasWhi18b,Taruya2019}.

\section{Redshift-Space Power Spectrum}
\label{sec:pk_rsd}

\edit{We now proceed to write down the power spectrum at one loop in Lagrangian perturbation theory \cite{Mat08b,CLPT,Wan14,Whi14,VlaCasWhi16,VlahWhite19,ChenVlahWhite20}. From Equation~\ref{eqn:deltam} and its counterpart for biased tracers we have that the galaxy autospectrum is given by
\begin{equation*}
    P(\bk) = \dtq  e^{i\bk \cdot \bq} \avg{e^{i\bk \cdot \Delta} F(\bq_1) F(\bq_2)}_{\bq = \bq_1 - \bq_2}
\end{equation*}
where we have defined the pairwise Lagrangian displacement $\Delta_i = \bPsi_i(\bq_1) - \bPsi_i(\bq_2)$. Setting $F = 1$, for matter, the bracketed average can expressed using the cumulant theorem as \cite{CLPT}
\begin{equation*}
    \ln \avg{e^{i\bk\cdot\Delta}} = -\half k_i k_j A_{ij} - \frac{i}{6} k_i k_j k_k W_{ijk} + ...,
\end{equation*}
where we have defined the cumulants of the pairwise displacements as $A_{ij} = \avg{\Delta_i \Delta_j}_c$ and $W_{ijk} = \avg{\Delta_i \Delta_j \Delta_k}_c$. For biased tracers one simply needs to compute cumulants with sources like $J(\bq) \delta(\bq)$ added to the exponent and take functional derivatives; this produces terms like $U_i = \avg{\delta_0(\bq_1) \Delta_i}.$ The above calculations can be promoted to redshift space by promoting the displacements to redshift space ($\Delta^s$) order-by-order as in Equation~\ref{eqn:PsiRSD}.
}

\edit{From the above, the one-loop} galaxy autospectrum in redshift space is given in LPT by \cite{Mat08b,CLPT,Wan14,Whi14,VlaCasWhi16,VlahWhite19,ChenVlahWhite20}
\begin{align}
    P_s(\bk)
    &= \dtq e^{i\bk \cdot \bq}\ e^{-\frac{1}{2}k_ik_j A^{s,<}_{ij}} \Big\{ 1  - \half k_i k_j A^{s,>}_{ij} + \frac{1}{8} k_i k_j k_k k_l A^{s,>}_{ij} A^{s,>}_{kl}\nonumber \\
    &- \half k_i k_j A^{s,\rm loop}_{ij} + \frac{i}{6}k_i k_j k_k W^s_{ijk} \nonumber \\
    & +2 i b_1 k_i (1 - \half k_i k_j A^{s,>}_{ij}) U^s_i - b_1 k_i k_j A^{s,10}_{ij} \nonumber \\
    &+ b_1^2 (1 - \half k_i k_j A^{s,>}_{ij}) \xi_{\rm lin}  + i b_1^2 k_i U^{s,11}_i - b_1^2 k_i k_j U^{s,\rm lin}_i U^{s,\rm lin}_j\nonumber \\
    &+ \frac{1}{2}b_2^2 \xi_{\rm lin}^2  + 2i b_1 b_2 \xi_{\rm lin} k_i U^{s,\rm lin}_i  - b_2 k_i k_j U^{s,\rm lin}_i U^{s,\rm lin}_j + i b_2 k_i U^{s,20}_i  \nonumber \\
    &+ b_s (-k_i k_j \Upsilon^s_{ij} + 2i k_i V^{s,10}_i) + 2 i k_i b_1 b_s V^{s,12}_i + b_2 b_s \chi + b_s^2 \zeta + 2 i b_3 k_i U^s_{b_3,i} + 2 b_1 b_3\theta + ... \Big\} \nonumber \\
    &+ k^2 (\alpha_0 + \alpha_2 \mu^2 + \alpha_4 \mu^4 + \alpha_6 \mu^6) P_{s,\rm Zel}(\bk)   + R_h^3 (1 + \sigma_2 k^2  \mu^2 + \sigma_4 k^4 \mu^4) .
\label{eqn:pks}
\end{align}
\edit{Here we have split the second cumulant $A_{ij}$ into long and short linear components and a loop component, keeping only the long-wavelength piece $A^{s,<}_{ij}$ exponentiated; we will comment on this further below. In addition to $A_{ij}$, $W_{ijk}$ and $U_i$ defined above, Equation~\ref{eqn:pks} contains additional correlators of pairwise Lagrangian displacements and higher-order bias operators like shear that are defined explicitly in refs.~\cite{CLPT,Whi14,VlaCasWhi16,ChenVlahWhite20}.} The last line of Equation~\ref{eqn:pks} includes counterterms ($\alpha_n$) and stochastic contributions ($\sigma_n$) proportional to the typical scale of halo/galaxy formation $R_h$.  \edit{These include what are traditionally referred to as the ``shot noise'' and ``finger of god'' (FoG) terms.  The small-scale sensitivities that give rise to these terms are described in detail\footnote{In particular, the advantages of this form for treating fingers of god (and redshift errors) are discussed in detail in \S\S 4.1.3, 4.2.3, 5.3 and Appendix C of ref.~\cite{ChenVlahWhite20}.} in ref.~\cite{ChenVlahWhite20}.}

While the correlators in Equation~\ref{eqn:pks} have been extensively described elsewhere, it is instructive to elucidate their general perturbative and angular structure with an example. Let us consider the displacement two-point function up to one-loop
\begin{equation}
    A_{ij} \equiv \avg{\Delta_i \Delta_j} = A^{\rm lin}_{ij} + A^{\rm loop}_{ij}
    \quad , \quad
    A^{\rm loop}_{ij} = A^{(22)}_{ij} + 2 A^{(13)}_{ij},
\end{equation}
which is given by a linear piece (lin) from contracting two first-order displacements and a one-loop piece from contracting two second-order displacements $(22)$ or one first and one third-order displacement each $(13)$. To go into redshift space, each of these pieces must be transformed separately --- this is because at each order in perturbation theory the translation to redshift space depends on $n$, such that for example
\begin{equation}
    A^{s,(13)}_{ij} = R^{(1)}_{in} R^{(3)}_{im} A^{(13)}_{nm}
    \quad\mathrm{but}\quad
    A^{s,(22)}_{ij} = R^{(2)}_{in} R^{(2)}_{im} A^{(22)}_{nm}
    \quad . \nonumber
\end{equation}
The $n$-dependence of these transformations encodes information about beyond-linear velocities within the RSD spectrum and is the primary complication in extending the treatment of RSD beyond the Zeldovich approximation, where all vectors transform via $\textbf{R} = \textbf{R}^{(1)}$ (see refs.~\cite{Whi14,VlahWhite19,ChenVlahWhite19}). A similar observation applies to all correlators with vector indices in Equation~\ref{eqn:pks}.

Finally, let us comment on our resummation of the linear piece of $A_{ij}$. A salient feature of cosmologies like $\Lambda$CDM is that large-scale displacements produce nonlinear damping of spatially localized features in the power spectrum such as baryon acoustic oscillations (BAO) that cannot be captured simply with an order-by-order expansion in the linear initial conditions \cite{Bha96,ESW07,Mat08a,Mat08b,Crocce08,PWC09,Noh09,CLPT,TasZal12,McCSza12,White14,SenZal15,Schmittfull15,Baldauf15,Vlah16,McQuinn16,Blas16,Seo16}. This is because, while the dynamics on these large scales are essentially linear, the size of these displacements on BAO scales can be large compared to wavenumbers where the BAO wiggles have support. As such, the effects of these displacements must be manually resummed in order-by-order expansions such as Eulerian perturbation theory (EPT). On the other hand, within LPT the exponential in Equation~\ref{eqn:deltam} and the cumulant theorem for a Gaussian variable
\begin{equation}
\label{eq:exp_damp}
    \avg{e^{i \bk \cdot \Delta^{(1)}}} = e^{-\half k_i k_j A^{(11)}_{ij}}
\end{equation}
suggests a natural resummation scheme wherein the linear displacements are kept exponentiated. These exponentiated displacements should be resummed only up to an IR scale $k_{\rm IR}$; we will follow the convention in ref.~\cite{ChenVlahWhite20} and perform this split by an exponential cutoff $\exp[-\half(k/k_{\rm IR})^2]$ in the $A_{ij}$ integral for the ``less than'' displacements kept resummed, $A^{s,<}_{ij}$ in Equation \ref{eqn:pks}, and the ``greater than'' displacements (defined with $1-\exp[-\half(k/k_{\rm IR})^2]$ in the integral), $A^{s,>}_{ij}$, which are expanded to second order in the curly brackets in Equation \ref{eqn:pks}. Within $\Lambda$CDM-like cosmologies, $A_{ij}$ is close to saturated on BAO scales and it might be expected that an IR cutoff should make only small differences in the final theory prediction.  This is true for the density statistics \cite{White14,McQuinn16,VlaCasWhi16,Baldauf15,Ivanov18}.  However, ref.~\cite{ChenVlahWhite20} showed that the higher-order velocity statistics that enter into RSD are especially sensitive to IR resummation in both broadband and BAO wiggles. This is discussed in further detail in Appendix~\ref{app:kir}.
We conclude this section by noting that, contrary to EPT where the a posteriori IR-resummation has been implemented only for equal time correlators, the direct evaluation of the LPT integrals presented in this work automatically evaluates the power spectrum at unequal times. In this case, the bulk displacement contributions do not cancel exactly in the exponent in Eq.\eqref{eq:exp_damp} leading to rapid suppression and decorrelation of unequal time correlators. This has recently been considered \cite{Chisari19} in the context of weak lensing analyses, that all involve unequal time correlators.  On the same topic, this should also clarify some recent concerns raised in ref.~\cite{delaBella:2020rpq} about the use of perturbation theory for unequal time correlators.

\section{Numerical Implementations}
\label{sec:numeric}

\begin{figure}
    \centering
    \includegraphics[width=\textwidth]{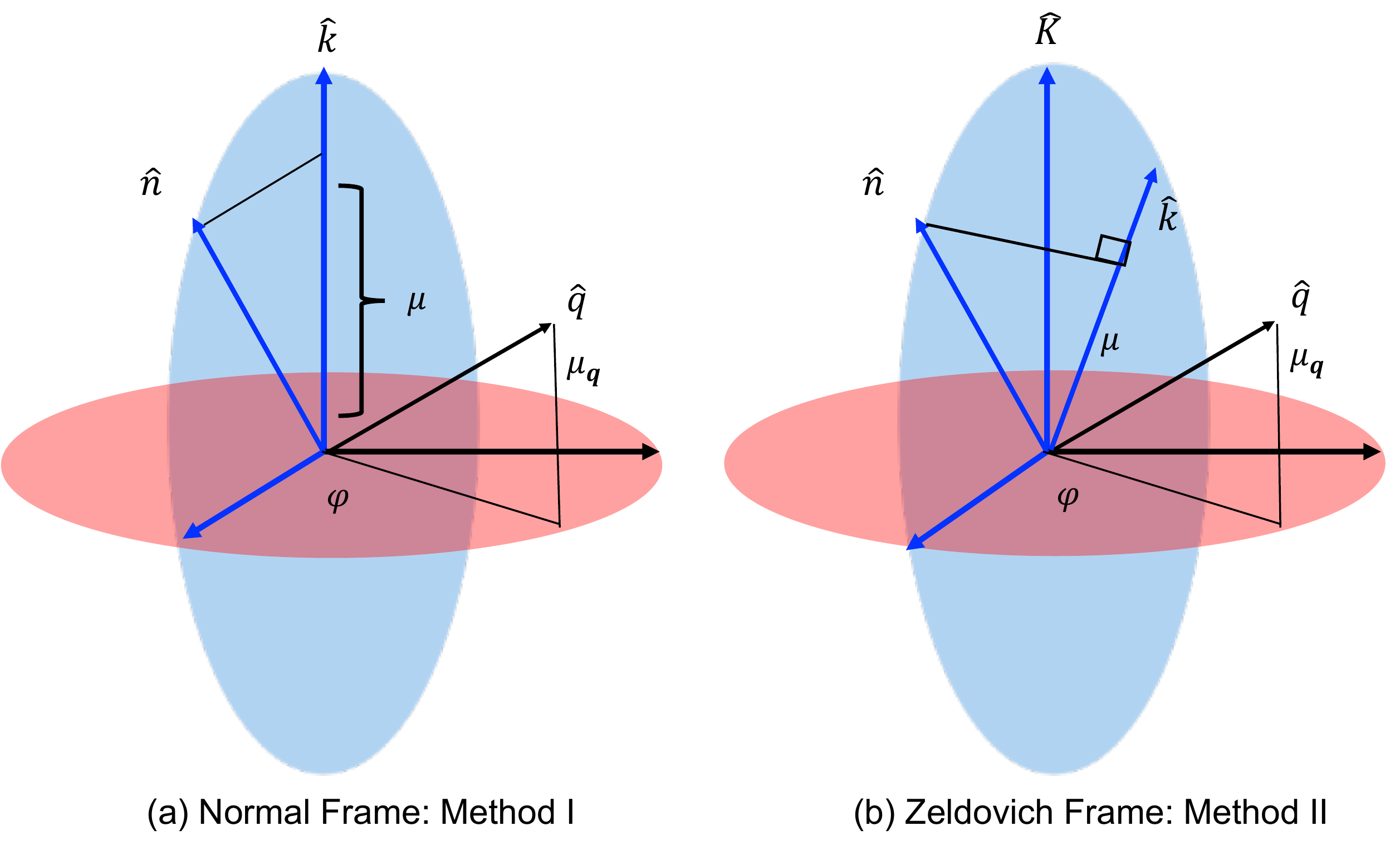}
    \caption{(a) Geometry of the vector and tensor quantities in the integral Eq.~\ref{eqn:pks}. In the absence of redshift-space distortions ($\hn$ dependence) the integral over $\bq$ is azimuthally symmetric; with RSD, a $\phi$ dependence occurs since $\hk$ and $\hn$ lie on a preferred plane. (b) In Method II, vectors are boosted into the ``Zeldovich'' frame where projections along the line of sight are amplified by the linear growth rate $f$ and the zenith is redefined to be the thus-boosted $K_i = R_{ij} k_j$. In both frames, $\hn$, $\hk$ and $\hat{K}$ are coplanar as shown in blue. }
    \label{fig:frames}
\end{figure}

The primary challenge in evaluating the integral in Equation~\ref{eqn:pks} lies in the angular dependence due to the three vectors, $\bq$, $\bk$ and $\hn$, that enter the calculation (Fig.~\ref{fig:frames}a). By symmetry, each of the tensor-indexed Lagrangian-space correlators in Equation~\ref{eqn:pks} can be decomposed into components multiplying products of $\hq$ and the Kronecker delta symbol; for example, we can write $A_{ij}(\bq) = X(q) \delta_{ij} + Y(q) \hq_i \hq_j$ \cite{CLPT}. In real space, where the angular dependence is due only to \edit{$\bk$ and $\bq$}, it is customary to proceed by defining a coordinate system wherein $\bk$ points towards the zenith such that the integrand has azimuthal symmetry. The resulting dependence on $\mu_{\bq} = \hk \cdot \hq$, where the subscript is meant to distinguish it from the familiar LOS angle $\mu = \hat{k}\cdot\hat{n}$, can then be recast into infinite sums of spherical Bessel functions using, for example, the identity
\begin{equation}
    \frac{1}{2} \int d\mu_\bq\ e^{iA \mu_\bq - \frac{1}{2} B \mu_\bq^2} = e^{-B/2} \sum_{n = 0} \Big( \frac{B}{A} \Big)^n j_n(A)
\end{equation}
and its derivatives, with the resulting integrals in $q$ efficiently computed using the FFTLog algorithm \cite{Ham00, VlaSelBal15}. In the particular case of LPT, we have $A = kq$ and $B = k^2 Y^{<}(q)$, such that in the Zeldovich power spectrum (second line of Equation~\ref{eqn:pks} in the $k_{\rm IR} = \infty$ limit) is
\begin{equation}
    P_{\rm Zel}(k) = 4 \pi \sum_{n=0}^\infty \int dq\,q^2 \ e^{-\half k^2 (X + Y)} \Big( \frac{kY}{q} \Big)^n j_n(kq)
\end{equation}
in real space. This form of the integral, 
suitable for using fast Hankel transforms, 
generalizes for higher loop terms, 
as well as when we go to redshift space, 
as we show in the rest of this section. 
Note that the computation calls for 
one transform for each $k$ value. 
The expansion converges quickly for the 
$k$ values of our interest and it is typically 
sufficient to keep only the $n<10$ terms in the
sum above. Moreover, for higher $n$ terms the
Limber approximation \cite{Limber:1953,LoVerde:2008} 
can be used 
\begin{equation}
j_{\ell}(kq) 
\approx 
\sqrt{\frac{\pi}{2\ell+1}}\ \delta^D\left( k q - \ell -\frac{1}{2} \right) , ~~{\rm when}~~ \ell \to \infty,
\end{equation}
which provides an accurate approximation for the integral above when used for $n>3$. 

In redshift space, the azimuthal symmetry is broken by the line-of-sight dependence, as shown in Figure~\ref{fig:frames}a, which selects a preferred plane containing $\hk$ and $\hn$ (blue). Below, we outline one method to efficiently perform the integral in redshift space. Following previous work \cite{VlahWhite19} we will call it Method II. Our development extends the Zeldovich calculations for matter and biased tracers in refs.~\cite{TayHam96,VlahWhite19,ChenVlahWhite19} to one-loop order.
Method II relies on an active transformation of the wavevector $\bk$ into a frame more conducive to evaluating the integral in Equation~\ref{eqn:pks}. Of course, it is also possible to directly evaluate the integral within the original frame --- this is the strategy of Method I. This alternative method is described for the interested reader in Appendix~\ref{app:mi}.

We now outline the rough strategy for Method II. In our expressions $\bPsi$ enters only in the combination $\bk\cdot\bPsi$.  Thus instead of transforming all displacements into redshift space via $\bPsi^{(n)}\to\textbf{R}^{(n)}\bPsi^{(n)}$ we can instead passively transform the wave vectors, multiplying them by $\textbf{R}^T=\mathbf{R}$.  In particular we shall apply the linear theory transformation $\textbf{R}^{(1)}$ to $\bk$ to yield the vector $K_i = R_{ij}^{(1)} k_j$. We will deal with the non-linear contributions to $\bPsi$ below. In this new ``Zeldovich'' frame, shown in Figure~\ref{fig:frames}b, the zenith direction is set to be $\hat{K}$, and we redefine $\mu_{\bq} = \hat{K} \cdot \hq$. Thus $\mathbf{K} \cdot \hat{n} = k \mu (1+f)$ and $K^2 = k^2[1 + f(2+f)\mu^2]$. Note that $\hk$, $\hat{K}$ and $\hn$ are coplanar.

Let us begin by reviewing the angular structure of this coordinate choice.  We have
\begin{align}
    &\hat{n} \cdot \hat{q} = A(\mu) \mu_{\bq} + B(\mu) \sqrt{1-\mu_{\bq}^2}  \,\cos\phi, \quad  \nonumber \\
    &\bk \cdot \bq = kq \Big(c(\mu)\mu_{\bq} - s(\mu)\sqrt{1-\mu_{\bq}^2}\cos\phi \Big),
    \label{eqn:MII_dots}
\end{align}
with the definitions
\begin{align}
    A(\mu) &= \frac{ \mu (1+f)}{\sqrt{1+f(2+f)\mu^2}} \quad , \quad B(\mu) = \sqrt{\frac{1 - \mu^2}{1 + f(2+f)\mu^2}} \nonumber \\
    c(\mu) &= \frac{1+f\mu^2}{\sqrt{1+f(2+f)\mu^2}} \quad , \quad s(\mu) = \frac{f\mu\sqrt{1-\mu^2}}{\sqrt{1+f(2+f)\mu^2}} \quad .
\end{align}
Note the square root in the denominators is simply $K/k$.  That the azimuthal dependence always multiplies the sine, $\sqrt{1-\mu_{\bq}^2}$, will prove a particular convenience in this frame.

In terms of the above, the Zeldovich matter power spectrum can be succintly expressed as
\begin{align*}
     P_s(\bk) &= \int dq\ d\mu_{\bq}\ q^2\ e^{ikqc\mu_{\bq}- \frac{1}{2}K^2(X+Y\mu_{\bq}^2)}\ \Bigg(\int d\phi\ e^{-ikqs\sqrt{1-\mu_{\bq}^2} \cos\phi} \Bigg)\  \nonumber
\end{align*}
In ref.~\cite{VlahWhite19} this integral was shown to be expressable in terms of Bessel functions via the identity
\begin{equation}
    I(A,B,C) = \int d\mu_{\bq}\ d\phi\ e^{-iC\sqrt{1-\mu_{\bq}^2}\cos\phi + iA\mu_{\bq} + B\mu_{\bq}^2} = 4\pi e^B \sum_{\ell=0}^\infty \Big( \frac{-2}{\rho} \Big)^\ell \tilde{G}^{(0)}_{0,\ell}(A,B,\rho) j_\ell(\rho)
    \label{eqn:int_mii}
\end{equation}
by substituting $A = kqc$, $B = -\frac{1}{2}K^2Y$ and $C = kqs$. The exact form of the the kernel $\tilde{G}^{(0)}_{0,\ell}$ is given in Appendix~\ref{ssec:g0_defs}. Building on top of this, any bias contribution involving only the linear (Zeldovich) displacement (e.g.\ the linear bias term $i k_i U^{\rm lin}_i$), simply acquires powers of $\mu_{\bq}$ and $K$ (e.g.\ $i K\mu_{\bq}$) that can be evaluated as derivatives of the above with respect to $A$, since correlators in Lagrangian space are always decomposable into $\delta_{ij}$ and tensor products of $\hq_i$ \cite{ChenVlahWhite19}.

The simple Zeldovich angular structure above is, however, broken by the inclusion of higher-order displacements. This is because these displacements get boosted along the line of sight by more than linear theory when going to redshift space.  With $\mathbf{R}=\mathbf{R}^{(1)}$
\begin{equation}
    \dot{\bPsi}^{(n)} = \vb{R}^{(n)} \bPsi^{(n)} = (\vb{R} + (n-1) f\ \hn \otimes \hn) \bPsi^{(n)}.
\end{equation}
In the spirit of the above calculations we can dot the matrix into the wavevectors and take
\begin{equation}
    k_i \rightarrow K_i + f (n-1) k_{\parallel,i} \quad , \quad k_\parallel = (k\mu) \hn.
\end{equation}
Dotting the transformed wavevector with Lagrangian correlators thus simply requires additional powers of $\hn \cdot \hq$, which conveniently translates into powers of $\mu_{\bq}$ and $\sqrt{1-\mu_{\bq}^2}\cos\phi$, i.e.\ the coefficients multiplying $A$ and $C$ in Equation~\ref{eqn:int_mii}. We thus see that any contribution to the power spectrum can be evaluated via mixed ($A,C$) derivatives of Equation~\ref{eqn:int_mii}. We refer the reader to Appendix~\ref{app:mii} for further details and an example application to the one-loop matter power spectrum.

\section{Results}
\label{sec:results}

Having laid out how the one-loop power spectrum can be efficiently computed within fully-resummed one-loop LPT, our goal in this section is to validate our model against N-body data. In addition, we compare the performance of our model with previous models such as the Gaussian streaming model (GSM; \cite{Pee80,Fis95,ReiWhi11,Reid12,Wan14,VlaCasWhi16}) and moment expansion (MOME; \cite{VlahWhite19,ChenVlahWhite20}) in LPT and resummed Eulerian perturbation theory (REPT) in both Fourier and configuration space.

\subsection{Comparison to N-body}
\label{sec:anl_sims}

\begin{figure}
    \centering
    \includegraphics[width=\textwidth]{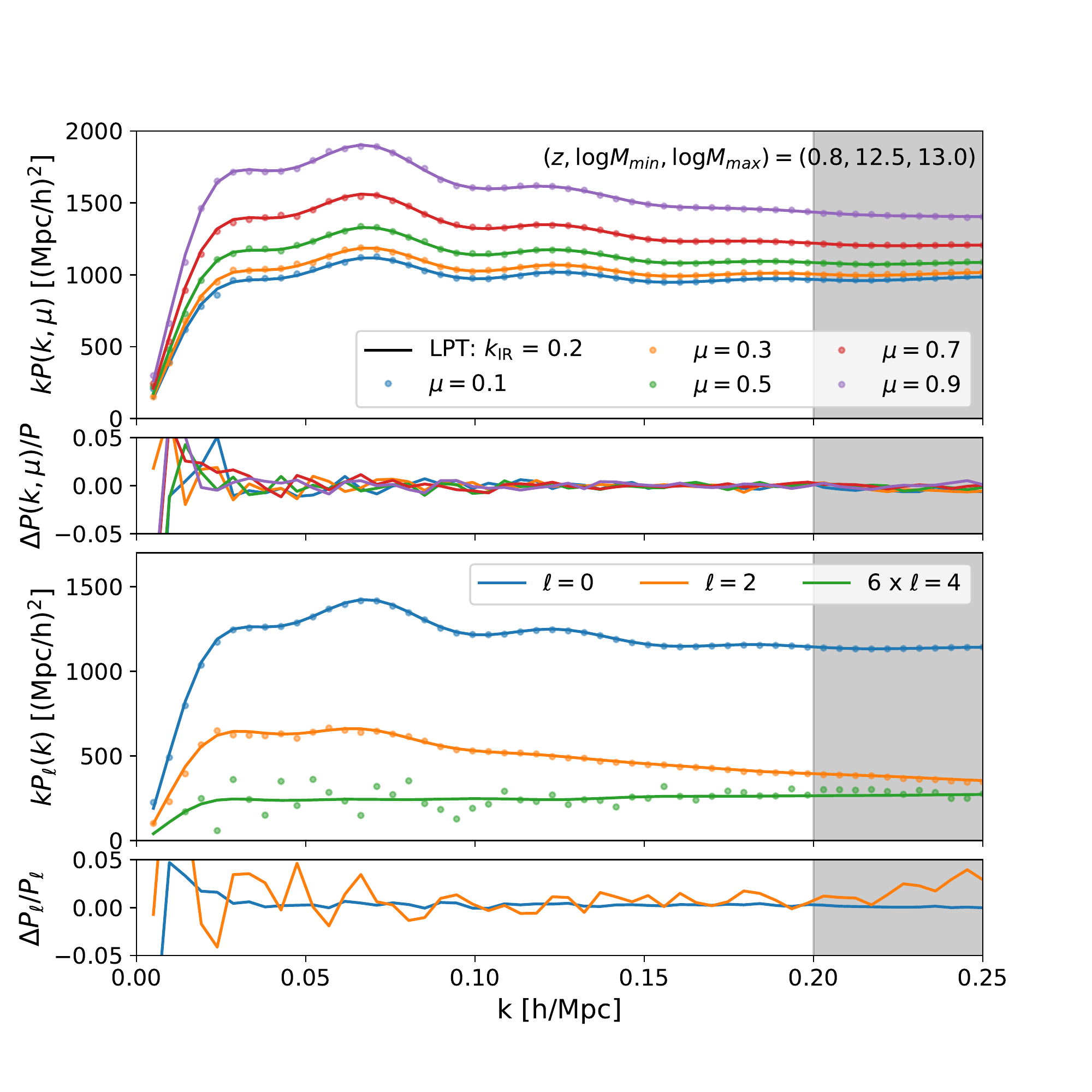}
    \caption{Fits to the redshift-space power spectrum wedges (top) and multipoles (bottom) of the fiducial halo sample with $10^{12.5} M_\odot < M < 10^{13.0} M_\odot$ at $z = 0.8.$ Both statistics were fit assuming Gaussian covariances using a consistent set of bias parameters and with linear displacements resummed up to $k_{\rm IR} = 0.2 \kMpc$. The fiducial LPT model gives an excellent fit to the anisotropic power spectrum inside the range of fit ($k < 0.2 \kMpc$) well within the few-percent systematics expected from the N-body data. Shaded regions indicate wavenumbers beyond the range of fit, with higher multipoles diverging faster from the data past this point.}
    \label{fig:lpt_fig}
\end{figure}

 For our main comparisons to N-body data we use halo catalogs from the simulations in ref.~\cite{Sunayama16} and employ {\sc NbodyKit} \cite{nbodykit} to compute redshift-space power spectrum wedges, multipoles and correlation function multipoles at $z = 0.8$. These simulations assume a $\Lambda$CDM cosmology with $\Omega_m = 0.2648$, $\Omega_b h^2 = 0.02258$, $h = 0.71$, $n_s = 0.963$ and $\sigma_8 = 0.8$. We adopt the mass bin $12.5 < \log(M/h^{-1}M_\odot) < 13.0$ as our fiducial sample but have checked that we get similar results for a higher mass bin as well as the mock galaxy sample described in ref.~\cite{ChenVlahWhite20}. \edit{Our fiducial sample has a number density of $\bar{n} = 0.53 \times 10^{-3} \icMpc$ and linear (Eulerian) bias of $b\approx 1.7$, making it slightly sparser but about 40\% more biased than the DESI ELG sample at $z = 0.85$ \cite{DESI}.} We have chosen these simulations due to their relatively large total volume (4 boxes with volume $[4 h^{-1}$Gpc$]^3$). With such a large volume the statistical errors on the two-point functions will necessarily be significantly smaller than galaxy surveys at comparable redshifts; however, we caution that the use of ``derated'' time steps in the running of these simulations may cause systematic errors on the few percent level, as discussed in further detail in refs.~\cite{VlaCasWhi16,ChenVlahWhite20}.  We have attempted to mitigate this effect by using only the high redshift catalog at $z = 0.8$.

Figure~\ref{fig:lpt_fig} compares our LPT model to the power spectrum wedges and multipoles of our fiducial halo sample. We fit for the wedges, $P(k,\mu)$, up to $k_{\rm max} = 0.2 \kMpc$ assuming Gaussian covariances.  We use the same parameters for the multipoles. We find bias parameters of order unity and the isotropic stochastic contribution $R_h^3$ comparable to the shot noise, noting that extending to higher (unperturbative) $k_{\rm max}$ tends to recover apparently good fits with anomalously large bias and effective parameters. Our model is in excellent agreement with the power spectrum wedges at the scales shown, differing from the data at levels comparable to their statistical uncertainty, with qualitatively similar behavior in the multipoles, though the anisotropic contributions ($\ell > 0$) diverge faster than the monopoles as expected due to the enhanced nonlinearity of halo velocities.

In Figure~\ref{fig:lpt_fig}, as well as throughout the main body of this work, we have adopted the fiducial choice of infrared cutoff $k_{\rm IR} = 0.2 \kMpc$. As discussed in Section~\ref{sec:pk_rsd}, compared to density statistics the velocity statistics' underlying redshift-space distortions have broadband shapes that are especially sensitive to the choice of infared cutoff. For example, as shown in ref.~\cite{ChenVlahWhite20} the monopole and quadrupole of the second moment of the pairwise velocity, responsible for contributions to the power spectrum proportional to the growth rate ($f$) squared, respectively have broadband shapes better captured by large and small $k_{\rm IR}$. One might thus hope to find an intermediate regime wherein both statistics are reasonably captured, and indeed we find that the choice $k_{\rm IR} = 0.2 \kMpc$ reproduces the hexadecapole better than either the fully-exponentiated limit ($k_{\rm IR} = \infty$) or $k_{\rm IR} = 0.$ In principle, the spirit of perturbation theory should demand that the expanded displacements $k^2 \Sigma^2_{>}$ be small\edit{\footnote{Here we define
\begin{equation}
    \Sigma^{2}_{<} = \frac{2}{3} \int \frac{dk}{2\pi^2} P_{\rm lin}(k)\ e^{-(k/k_{\rm IR})^2}, \quad \Sigma^{2}_{>} = \frac{2}{3} \int \frac{dk}{2\pi^2} P_{\rm lin}(k)\ (1 - e^{-(k/k_{\rm IR})^2}),
\end{equation}
such that the sum $\Sigma^2 = \Sigma^2_{<} + \Sigma^2_{>}$ is the mean square pairwise displacement of two distant points in the Zeldovich approximation.
}} while the exponentiated ones kept manageable; for $k < 0.2 \kMpc$ this is satisfied by our choice, though given that the \textit{total} Zeldovich displacement for the fiducial cosmology at $z = 0.8$ is $\Sigma^{-1} \approx 0.2 \kMpc$ this is relatively insensitive to the choice of IR cutoff. \edit{Moreover, while differences exist towards high $k$ and $\mu$, we find that in general the small-scale differences between the theory's predictions for reasonable values of $k_{\rm IR}$ can largely be absorbed by the effective parameters of the theory; f}urther discussion of the interplay between $k_{\rm IR}$ choice and our model's predictions can be found in Appendix~\ref{app:kir}; we intend to return to this topic in greater depth in future work.

\subsection{Comparison to Other Models in Fourier and Configuration Space}

\begin{figure}
    \centering
    \includegraphics[width=0.9\textwidth]{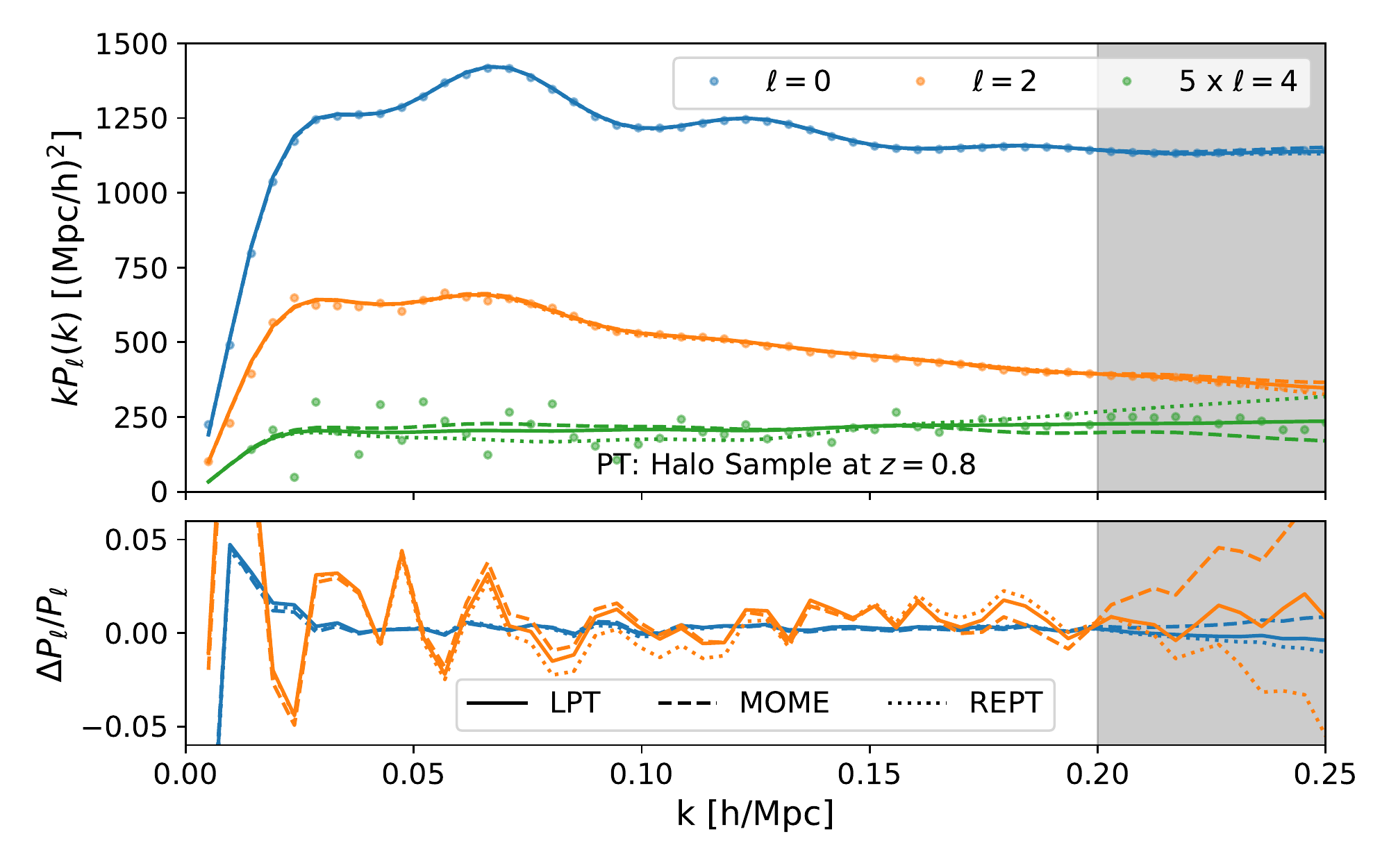}
    \caption{Redshift-space power spectrum multipoles of the fiducial halo sample fit using three effective theory models: the fiducial LPT model, the Lagrangian moment expansion and resummed Eulerian perturbation theory. All three models are fit as in Figure~\ref{fig:lpt_fig} and are in excellent quantitative agreement with the N-body data. The three models differ slightly in their prediction for the hexadecapole broadband; we have explicitly tuned our LPT IR resummation scheme to provide a good match to the data, though we note the relatively large statistical uncertainty in the hexadecapole.}
    \label{fig:pt_pells}
\end{figure}

\begin{figure}
    \centering
    \includegraphics[width=\textwidth]{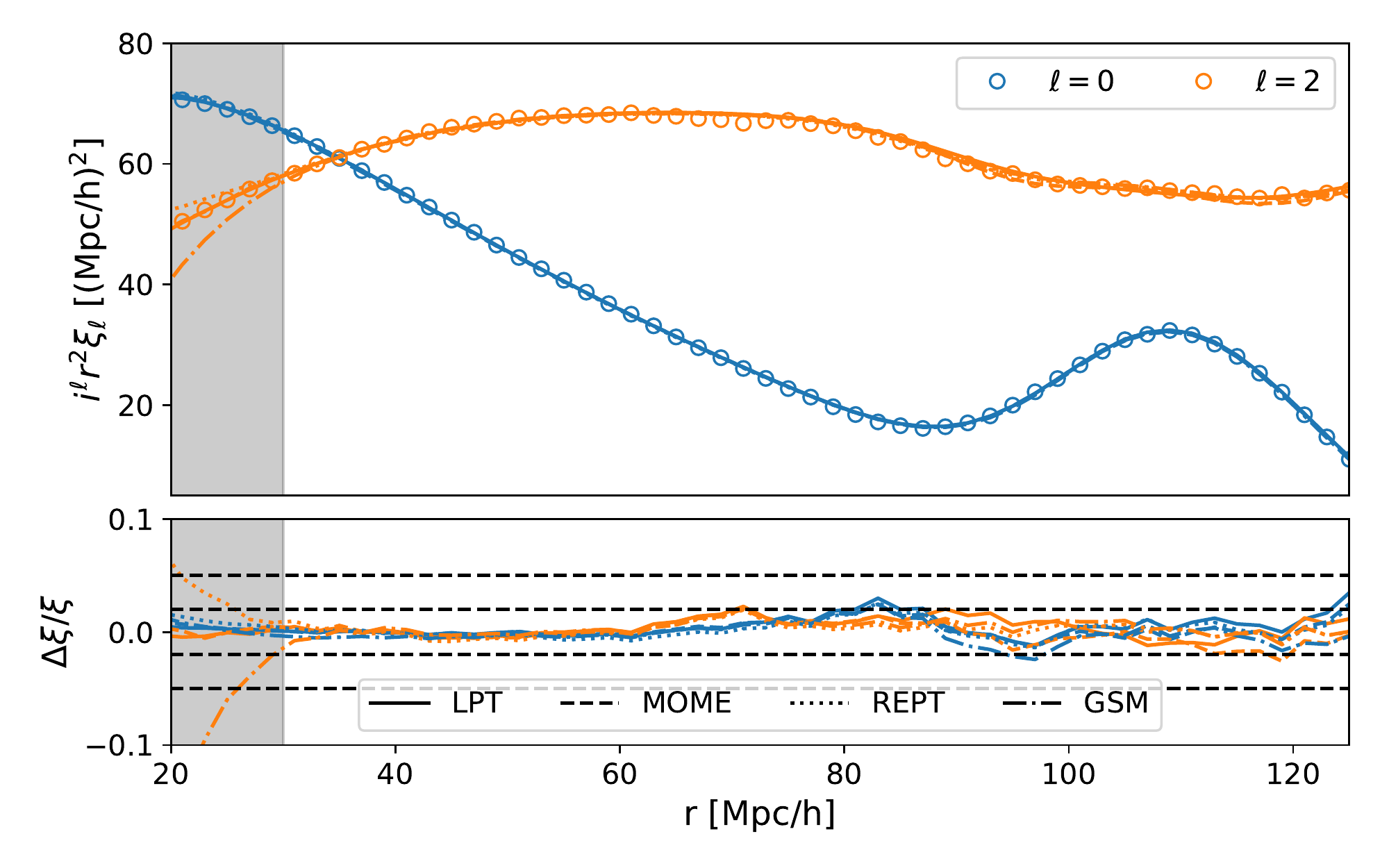}
    \caption{Configuration space correlation function multipoles predicted by our LPT model, the Lagrangian moment expansion, Eulerian perturbation theory and the Gaussian streaming model compared to N-body data. Each of the models are in good agreement with the data within the few-percent systematic uncertainties expected of the simulations, though we note that they all slightly overshoot the dip around $80\ h^{-1}$ Mpc by around two percent. Note that due to the high degree of similarity between the theory predictions many of the lines lie on top of each other even in the fractional residuals in the bottom panel, particularly when comparing LPT (solid) and MOME (dashed). Black dashed lines in the lower panels indicate 2 and 5 percent errors.
    }
    \label{fig:halo_xiells}
\end{figure}

The main difference between the model presented in this work and previous effective-theory models of the redshift-space galaxy two-point function lies in the IR-resummation procedure. Existing LPT formulations typically incorporate bulk velocities either via streaming model resummations \cite{VlaCasWhi16,VlahWhite19,ChenVlahWhite20} or direct expansions of velocity statistics \cite{VlahWhite19, ChenVlahWhite20}. 
While these approaches have been shown to be sufficiently accurate to model redshift-space distortions in a variety of contexts, they have nonetheless exhibited a number of shortcomings. The Gaussian streaming model allows for a partial resummation. However, the resummation procedure calls for nonlinear mapping of all the loop contributions, even those that require counterterms in order to regularize the UV dependence. Even though such mapping could in principle be restricted to only long wavelength contributions, the model also exhibits a somewhat cumbersome structure in Fourier space. The moment-expansion approach, on the other hand, relies on an explicit expansion in the velocity moments. However, in a similar way to the long wavelength displacement contributions, long wavelength velocity contributions also affect the BAO feature in a manner that then prompts the additional resummation of these contributions.  In \edit{the Eulerian approach} this can be done a posteriori using an ad hoc, wiggle-no-wiggle splitting of the power spectrum and resumming only the contributions related to the BAO feature \edit{(REPT)}. Such a procedure could also be performed for long wavelength velocity contributions in MOME, but this was not done in ref.~\cite{ChenVlahWhite20} where only long wavelength displacements were resummed in the LPT manner. We outline this resummation of long wavelength velocities in Appendix \ref{app:kir}. While the MOME approach has been shown to give excellent predictions for the Fourier-space power spectrum, configuration space statistics (where the BAO focus is not merely few-per cent oscillations on top of the broadband) are expected to be more sensitive to the details of IR resummation and BAO damping.

More generally, IR resummation has also been extensively studied in the Eulerian context \cite{Baldauf15,Vlah16,Bla16,Ivanov18,ChenVlahWhite20b}. Most often, these rely on the wiggle-no-wiggle splitting procedure, separating the smooth and BAO components of the linear power spectrum (see, e.g.\  \cite{ESW07,Vlah16,ChenVlahWhite20b}). This procedure allows for a simplified treatment of the nonlinear effects of the BAO where typically only leading effects are captured, neglecting the more intricate structure captured by LPT. \edit{Nonetheless, the controlled approximations that enter into this form of IR resummation are generally subdominant to higher-order (two-loop) corrections that have been studied in e.g.\ refs.~\cite{Vlah16, Blas16, Ivanov20}.}  In addition to these EPT approaches, refs.\ \cite{Lewandowski2015a,Perko2016} take an intermediate approach. These rely on an LPT-like resummation procedure that tries to preserve the unresummed EPT broadband behaviour, thus effectively retaining an EPT-like perturbative structure. In this approach, the anisotropic part of the exponent in Eq.~\eqref{eq:exp_damp} is expanded while only the isotropic part is left resummed (see also Appendix B of ref.\ \cite{VlaWhiAvi15} for a more detailed connection between the two approaches).

Our goal in this subsection is to investigate how one-loop LPT with long wavelength velocity contributions fully resummed compares to the approaches mentioned above, focusing on the anisotropic redshift-space broadband and the BAO feature in configuration space.
Figure~\ref{fig:pt_pells} compares the Fourier-space multipoles predicted by our fiducial LPT framework to the LPT moment expansion (MOME) and one-loop resummed EPT (REPT). All three frameworks are fitted assuming Gaussian covariances up to $k_{\rm max} = 0.2 \kMpc$ in $P(k,\mu)$ as in the previous subsection. All three frameworks show excellent agreement with the data, with any disagreements, including inter-framework disagreements, well within the few percent systematic errors we expect from these simulations. The frameworks differ most in the the hexadecapole, with the pure LPT framework apparently a better fit to the broadband shape over the scales shown; this should be taken with a grain of salt, however, as the statistical errors are large and we specifically checked our IR resummation procedure for the LPT framework using these data. Similarly, in Figure~\ref{fig:halo_xiells} we fit the correlation function multipoles of the fiducial halo sample using the three frameworks above as well as the Gaussian streaming model (GSM). All four are in excellent agreement with regards to both the BAO feature and broadband shape at quasi-linear scales. Since the correlation function multipoles probe a slightly different combination of modes than their Fourier-space counterparts with a hard $k$ cut, we have adjusted the best-fit bias parameters ``by eye'' to yield a better fit at $r > 30\ h^{-1} \text{Mpc}$, though we note that directly transforming the previous Fourier-space results still yield theoretical predictions within the few-percent systematic errors expected from these simulations. Together, Figures~\ref{fig:pt_pells} and \ref{fig:halo_xiells} suggest that, despite differences in IR resummations schemes, existing effective-theory frameworks of the redshift-space two-point function offer similar levels of performance on pertubative scales. Finally, we note that, while in the above comparisons we have independently fit the bias parameters and EFT corrections of each model to most favorably evaluate the performance of each, their bias bases can in principle be mapped onto each other order-by-order; when the bias parameters are thus fixed, these models will tend to make slightly different predictions due to differences in resummed IR modes at higher order. We discuss these differences in Appendix~\ref{app:kir}.

\subsection{Cosmological Constraints using Blind Challenge Data}
\label{sec:challenge}

\begin{figure}
    \includegraphics[width=\textwidth]{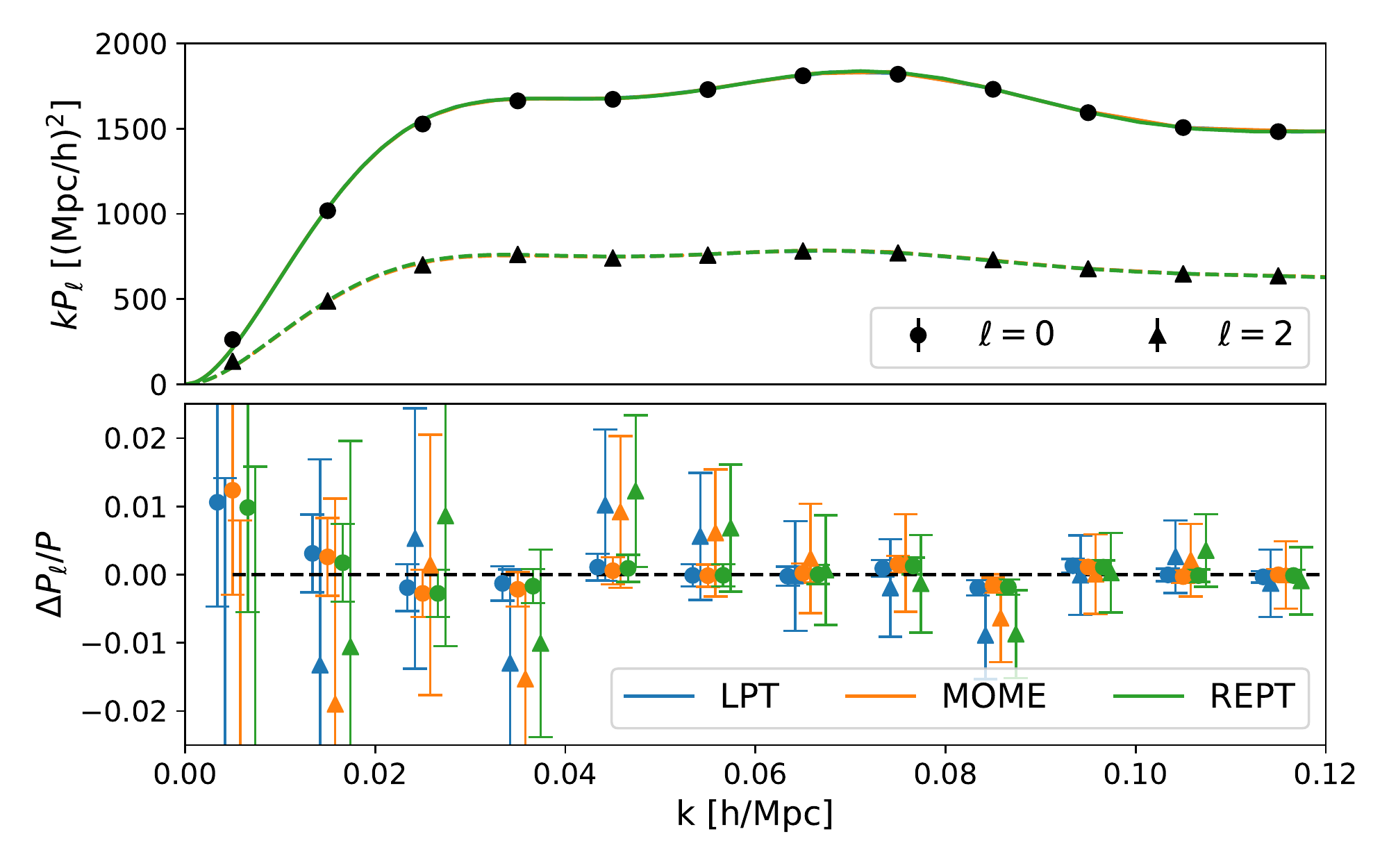}
    \caption{Multipoles of the blind challenge power spectrum along with the best fit one-loop LPT, MOME and REPT models. The top panels shows the (unbinned) theory curves along with the data. Both the error bars and theory differences are too small to see except in a few places. The lower panels show the fractional residuals of each (binned) theory curve, with each $k$ and $\ell$ bin separated by $0.0008 \kMpc$ for clarity of presentation.
    }
    \label{fig:blind_challenge_pk}
\end{figure}

\begin{figure}
    \begin{subfigure}{.49\textwidth}
    \centering
    \includegraphics[width=\linewidth]{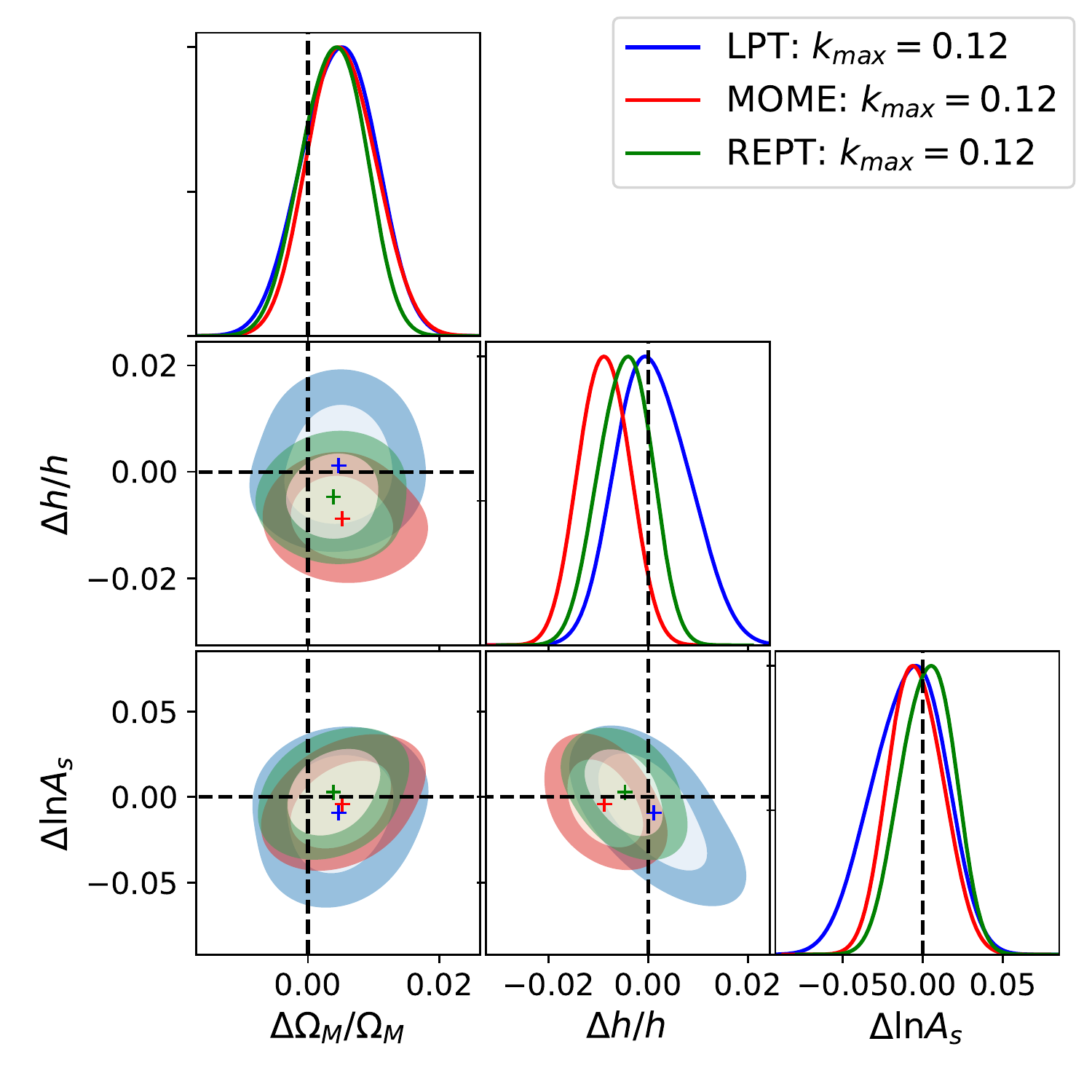}
    \end{subfigure}
    \begin{subfigure}{.49\textwidth}
    \centering
    \includegraphics[width=\linewidth]{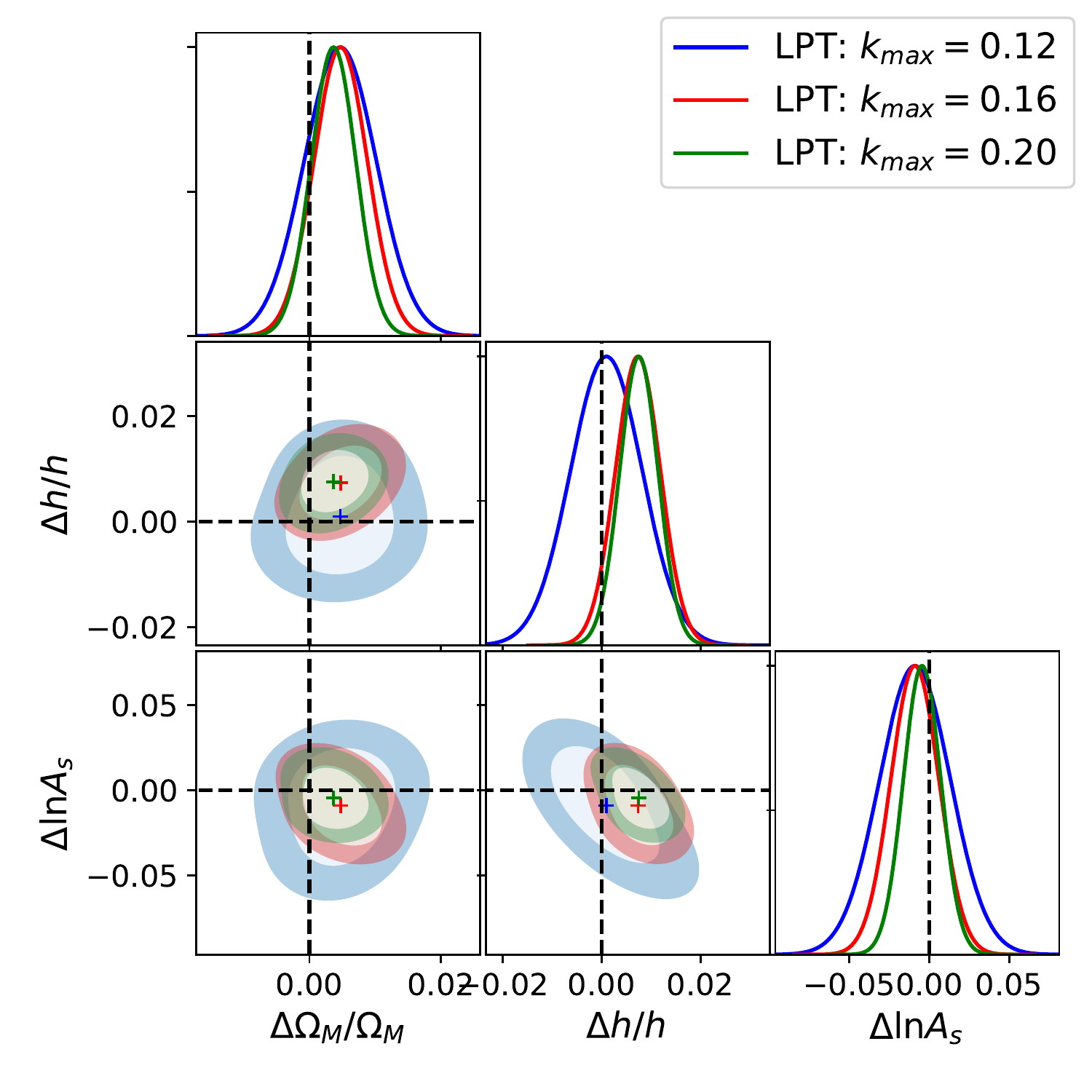}
    \end{subfigure}
    
    \caption{Histograms and two-dimensional contours for the three cosmological parameters in the blind challenge. (Left): Comparison of the LPT model in this paper to our previous submissions using the moment expansion (MOME) and resummed Eulerian perturbation theory (REPT), all at $k_{\rm max} = 0.12 \kMpc$. The LPT model performs competitively to existing models and indeed slightly improves upon the LPT-based MOME model's constraints on $h$. (Right): The LPT model constraints using three different scale cuts. All three scale cuts recover the truth on these parameters to within $2\,\sigma$.
    }
    \label{fig:blind_challenge}
\end{figure}

As a final test of our LPT model, we use it to model the ``blind challenge'' data described in ref.~\cite{Nishimichi2020}\footnote{The data and more information about the blind challenge can be found at \url{https://www2.yukawa.kyoto-u.ac.jp/~takahiro.nishimichi/data/PTchallenge/}.}. These are redshift-space power spectra for a BOSS-like HOD sample at $z\simeq 0.6$ constructed from ten N-body boxes each with sidelength $L = 3.84\,h^{-1}$Gpc sampled with $3072^3$ equal mass particles.  Since the total volume amounts to about 100 times the volume of the BOSS DR12 sample \cite{Alam17}, the statistical error associated with these data are expected to be far below any realizable galaxy survey at this redshift. These data were designed for a blind challenge wherein three cosmological parameters ($\Omega_M, h, \ln [10^{10} A_s]$) need to be fit assuming these tiny statistical errors while the baryon fraction, $f_b$, and spectral tilt, $n_s$, are fixed to the values used in the simulations. The challenge was designed to evaluate the performance of different PT approaches.  Any group wishing to enter the challenge submits their best fit cosmological parameters, without knowing the true ones, to Takahiro Nishimichi and collaborators. After submission one discovers if the model provides an unbiased estimate of the parameters.

We had previously submitted best fit parameters for MOME and REPT models that included scales up to $k_{\rm max} = 0.12 \,\kMpc$, and in both cases the models ``passed'': our REPT submission yielded means for the cosmological parameters well within $1\,\sigma$ of the truth, while MOME yielded $1.1\,\sigma$ and $1.8\,\sigma$ deviations for $\Omega_m$ and $h$, both well within errors expected for realistic galaxy surveys\footnote{Indeed, MOME also yields errors below $1\,\sigma$ for $k_{\rm max} = 0.14 \kMpc$, though we did not know this prior to submission and unblinding.} and possibly consistent with fluctuations in the challenge data themselves.
We have repeated the same exercise with the direct LPT model discussed in this work, using the same set of parameters\footnote{These are $b_1, b_2, b_s, \alpha_0, \alpha_2, R_h^3, \sigma_2$ in the Lagrangian basis and the equivalent set mapped onto the Eulerian basis. We have dropped the Lagrangian third-order bias because it is expected to be small and somewhat degenerate with other terms, $\alpha_{4,6}$ because we fit only up to the quadrupole and $\sigma_4$ because it was not necessary to fit the data at the scales we fit.}. As we have already participated in the challenge, we now know the true cosmological parameters.  However this should not affect the evaluation of our new LPT approach, since the analysis pipeline is the same one we adopted for MOME and REPT and we did not change the model in any way from that described in previous sections in order to participate in the challenge except to use the unblinded values as a seed in the MCMC to more quickly reach the maximum likelihood region. As in our previous submissions, uninformative priors were placed on all of the model parameters.

Figure \ref{fig:blind_challenge_pk} shows the measurements of the multipoles of the power spectrum along with the best fit one-loop LPT, MOME and EPT models. It's worth noticing that since we are also fitting for cosmological parameters, compared to the previous section where the linear power spectrum was held fixed and we varied only the bias parameters, the model has to include the Alcock-Paczynski (AP) effect\footnote{\edit{Specifically, we use Equations 42-46 of ref.\ \cite{Beu16}, though without the factors of $r_s$.}} \cite{APtest,Beu16}. Since this plot is just for visual comparison, we only show the best fit model with $k_{\rm max} = 0.12 \kMpc$. These data, produced using a different N-body code, halo finder and HOD prescription at a different redshift than the simulations in \S\ref{sec:anl_sims}, act as an additional test of the three PT models, and indeed the agreement between the models and with the data is remarkable up to the smallest scales included in the fit. All three models have $\chi^2/\text{dof} \approx 13 / (24 - 10)$, demonstrating good fits compared to their degrees of freedom.

Turning to the cosmological parameters, the left panel of Figure \ref{fig:blind_challenge} shows the $1\,\sigma$ and $2\,\sigma$ constraints obtained by fitting the monopole and quadrupole of $P(k)$ up to $k_{\rm max} = 0.12 \kMpc$ using our LPT, MOME and REPT models. This was the scale cut we chose in submitting results using MOME and REPT to the blind challenge, conservatively selected given the unusually low statistical uncertainty of the sample, as well as the main case analyzed in ref.~\cite{Nishimichi2020}.  
The LPT model performs slightly better than the other two, providing unbiased constraints on the three cosmological parameters.  In particular the bias in the Hubble constant, $h$, is reduced in LPT compared to both MOME and REPT. While the difference is less than $2\,\sigma$ and therefore well within the realm of possible statistical fluctuations in the N-body data, the improvement in our $\Omega_M$ and $h$ constraints, particularly relative to MOME in which bulk velocities are not fully resummed, suggests that our IR resummation scheme is correctly capturing the effects of large scale modes on both the BAO feature and broadband shape.

\begin{figure}
    \centering
    \includegraphics[width=\textwidth]{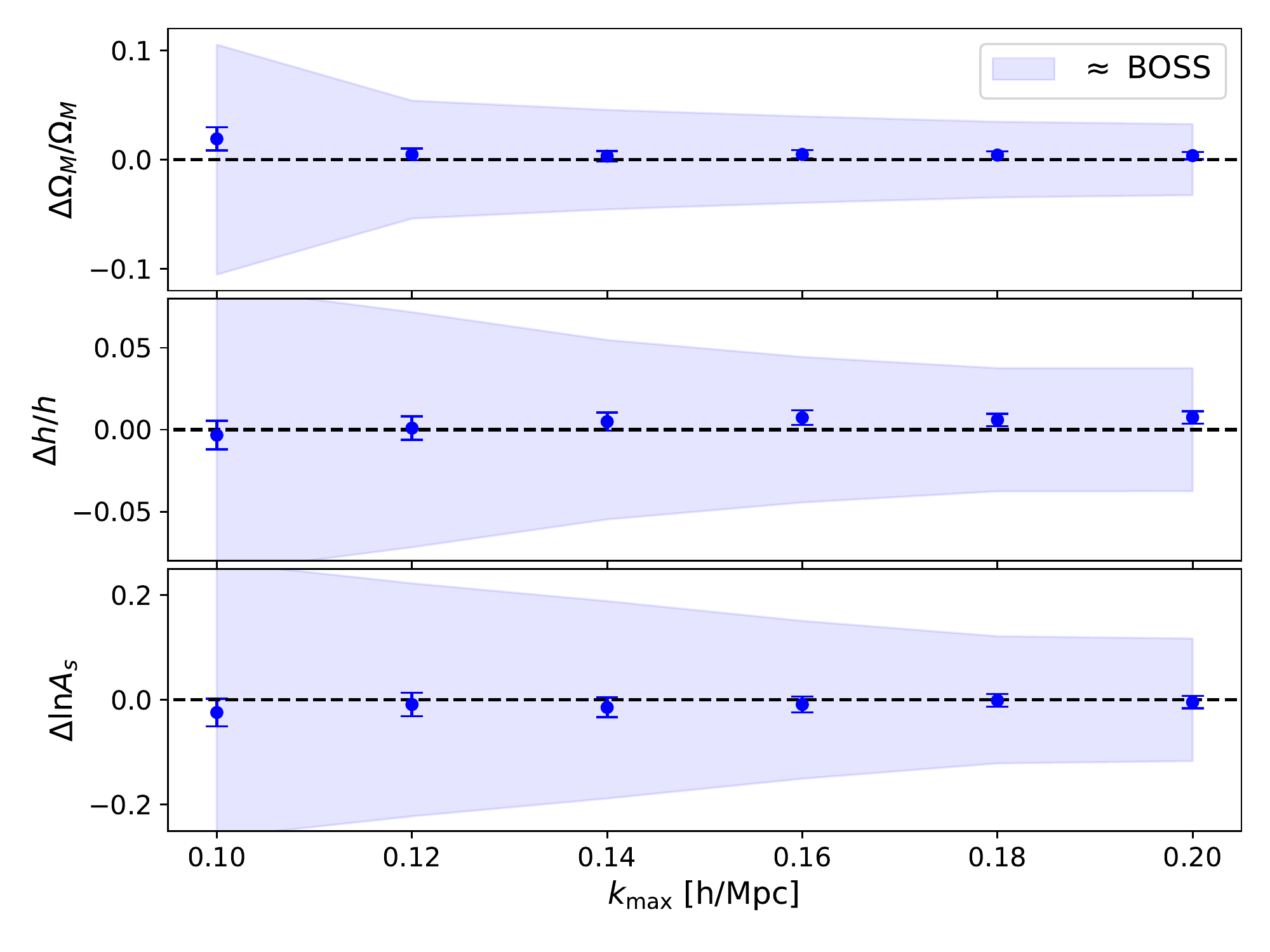}
    \caption{Parameter constraints using the LPT model for the blind challenge as a function of scale cut $k_{\rm max}$. The shaded blue region show errors scaled to a survey of $10 \times$ less volume, i.e. BOSS or a $\Delta z = 0.1$ slice of DESI at $z = 1.2$. All constraints shown are within $2\,\sigma$ of the truth, and within $1\,\sigma$ when $k_{\rm max} \leq 0.14 \kMpc$, well within error bars expected for future surveys as well as the realm of possible statistical fluctuations for this simulated sample.}
    \label{fig:kmaxs}
\end{figure}

The right panel in Figure \ref{fig:blind_challenge} shows the two-dimensional confidence intervals for different choices of $k_{\rm max} = 0.12$, 0.16, $0.20\kMpc$ for LPT. The constraints are within $2\,\sigma$ of the truth for each scale cut, but at $k_{\rm max} = 0.20 \kMpc$ the Hubble parameter $h$ has a mean very close to $2\,\sigma$ away from the truth while all parameters are well within $1\,\sigma$ at $0.12 \kMpc$, suggesting growing systematic bias at higher scale cuts where higher-order corrections are expected to play a more significant role. However, it should be noted that since these error bars are derived from the covariance of the sample itself --- within the Gaussian approximation no less --- the exhibited errors are well within the realm of statistical possibility and we cannot conclusively determine that any model is biased. These results are summarized in Figure~\ref{fig:kmaxs}, which shows the shift between the inferred and true bias parameters as a function of $k_{\rm max}$. The shaded regions indicate $10 \times$ the derived standard deviations, i.e.\  approximately the expected errors for a survey like BOSS, or a redshift slice of $\Delta z = 0.25$ at $z = 0.6$ for DESI. The LPT model presented in this work correctly recovers the underlying cosmology for all the scale cuts shown to well within the expected errors of surveys like BOSS and DESI.  We anticipate that the model would perform even better at higher redshift where the degree of non-linearity is smaller. We therefore conclude that, at least for the cosmological parameters probed in the challenge, our LPT model should provide an accurate tool for modeling RSD in upcoming surveys. 

\section{Conclusions}
\label{sec:conclude}

The anisotropic galaxy clustering observed by spectroscopic surveys probes density and velocity fields on large scales, enabling us to test the growth of structure in the quasilinear regime as predicted by General Relativity. In addition, the baryon acoustic oscillations in the galaxy clustering signal provide geometric information that constrain the cosmological expansion history. Perturbation theory is an ab initio approach with clear physical assumptions, and it is therefore the preferred tool for a  precise and rigorous mapping between cosmological parameters and the observed clustering signal, in the quest for a better understanding of the cosmological model and in the search for new physics.

The main purpose of this work was to further develop the modeling of the redshift-space two-point function within Lagrangian perturbation theory. Two critical aspects of any perturbation-theory model of the power spectrum or correlation function are the treatment of the nonlinear damping of the BAO signal due to large scale motions of galaxies and the inclusion of redshift-space distortions due to a degeneracy between the observed line-of-sight positions and peculiar velocities of galaxies. By expanding directly in displacements, Lagrangian perturbation theory naturally treats both phenomena within the same framework without relying on the wiggle-no-wiggle splitting procedure to damp the BAO oscillations. In particular, LPT exponentiates linear, or Zeldovich, displacements via the cumulant theorem, thereby resumming the long-wavelength (IR) modes primarily responsible for nonlinear BAO damping.

In this work, we extend the numerical techniques developed in refs.~\cite{VlahWhite19,ChenVlahWhite20} for the Zeldovich approximation to calculate the one-loop LPT power spectrum of biased tracers with both bulk displacements and velocities resummed. This is in contrast to previous efforts to model redshift-space distortions within effective LPT that focused on predicting velocity statistics for which only the displacements were exponentiated\footnote{Earlier work, e.g.\ ref.~\cite{CLPT}, performed a similar calculation as the one here for the correlation function only, but exponentiated the modes coming from the one-loop contributions as well.}. In Sections~\ref{sec:lpt} and \ref{sec:pk_rsd}, we outlined the effective formalism developed in those works and used them to write down the redshift-space power spectrum including third-order biasing, counterterms and stochastic contributions\edit{, of which the latter two in part play the role of ``finger-of-god'' terms in traditional RSD models.} Then, in Section~\ref{sec:numeric}, we developed the required numerical framework for the implementation of the model. We have publically released our implementation of Method II as part of the {\tt velocileptors} code.\footnote{https://github.com/sfschen/velocileptors} Additional details of these calculations can be found in Appendices \ref{app:mi} and \ref{app:mii}.

Finally, we compare the LPT model developed in this work to N-body data and previous models in Section~\ref{sec:results}. First, we fit the power spectrum of a halo sample drawn from a set of N-body simulations at $z = 0.8$ with masses $12.5 < \log(M/M_\odot) < 13.0$ assuming Gaussian covariances using our model in Figure~\ref{fig:lpt_fig}, finding excellent agreement in both the wedges and first three multipoles for a consistent set of bias and effective parameters. Our model has an extra degree of freedom in the IR cutoff $k_{\rm IR}$, which dictates the wavelength beyond which displacement modes are resummed. In ref.~\cite{ChenVlahWhite20} it was shown that the higher-order velocity statistics that enter into redshift-space distortions have broadband shapes that are especially sensitive to this choice, and indeed we find that a choice of $k_{\rm IR} = 0.2 \kMpc$, which lies between the broadband predictions of full-expanded LPT and EPT, gives the best match to the hexadecapole; this choice is explored further in Appendix~\ref{app:kir}, where we also show (Fig.~\ref{fig:xiell_kir}) that the configuration-space anisotropic BAO feature is remarkably robust to this hyperparameter. 

We then compare the performance of our LPT model with other existing effective-theory models. In Figure~\ref{fig:pt_pells}, we compare the aforementioned power spectrum predictions to the Lagrangian moment expansion (MOME) and resummed Eulerian perturbation theory, finding that all three can fit the data to within the expected systematic error of the simulations. Similar results for the configuration-space multipoles are shown in Figure~\ref{fig:halo_xiells}, where we also compare to the Lagrangian Gaussian streaming model.  As a last numerical test we checked whether the full one-loop LPT model presented in this work can recover unbiased cosmological parameters in a data analysis challenge using a different N-body code, halo finder, halo occupation distribution and redshift.  We find that LPT performs better than MOME and REPT at the reference scale cut chosen in ref.~\cite{Nishimichi2020}, and is able to recover the true cosmology with negligible biases up to $k_{\rm max} = 0.2\kMpc$ in a volume approximately one hundred times that of DESI or Euclid at the same redshift.  Results of this test are given in Figure \ref{fig:blind_challenge}.

Let us conclude by noting some possible extensions of our model and numerical implementation. There has been considerable recent interest in extending the bias expansion of galaxies to include anisotropic selection effects and, indeed, the redshift-space galaxy density can be decomposed into a generalized expansion of operators with LOS symmetry \cite{Hirata09,Desjacques18,Givans20,Obuljen20}. Since the angular dependencies of these operators will in general involve only tensor products of $\hk$, $\hn$ and $\hq$ their 2-point functions (with IR displacements resummed) should follow straightforwardly from our calculations.  Our calculations should also be straightforwardly extendable to modeling the reconstructed galaxy power spectrum at one-loop order, especially the ``Rec-Sym'' scheme \cite{ChenVlahWhite20} which features an identical structure, with the only difference being a larger set of terms involved. 
Concerning extensions of the $\Lambda$CDM model, perhaps the simplest one to implement is massive neutrino cosmologies. It is well known that halos and galaxies are biased tracers of the dark matter and baryon fluids only \cite{Villaescusa-Navarro2013,Castorina2013,Castorina2015,Villaescusa-Navarro2017,LoVerde2014,Munoz2018,Fidler2018}, which implies the bias expansion presented in Section \ref{sec:lpt} will still be valid with the trivial replacement of the total matter field with the dark matter plus baryon one.  Extra care should be taken with RSD since the growth rate is now scale dependent and Equation \ref{eqn:PsiRSD} is no longer valid \cite{Aviles:2020cax} (and similarly in modified gravity theories \cite{AviCer17,Valogiannis2019}). However given the smallness of neutrino masses, this complication is usually neglected when evaluating loop integrals.
We intend to return to these, admittedly more involved, calculations in future work.

\acknowledgments

We thank Jahmour Givans, Marko Simonovi\' c and the anonymous referee for helping us clarify various points in the draft. S.C.\ is supported by the National Science Foundation Graduate Research Fellowship (Grant No.~DGE 1106400) and by the UC Berkeley Theoretical Astrophysics Center Astronomy and Astrophysics Graduate Fellowship.
M.W.\ is supported by the U.S. Department of Energy and the NSF.
Z.V.\ is supported by the Kavli Foundation.
This research has made use of NASA's Astrophysics Data System and the arXiv preprint server.
This research used resources of the National Energy Research Scientific Computing Center (NERSC), a U.S. Department of Energy Office of Science User Facility operated under Contract No. DE-AC02-05CH11231.

\appendix

\section{Infrared Resummation and the Broadband}
\label{app:kir}

\begin{figure}
    \centering
    \includegraphics[width=\textwidth]{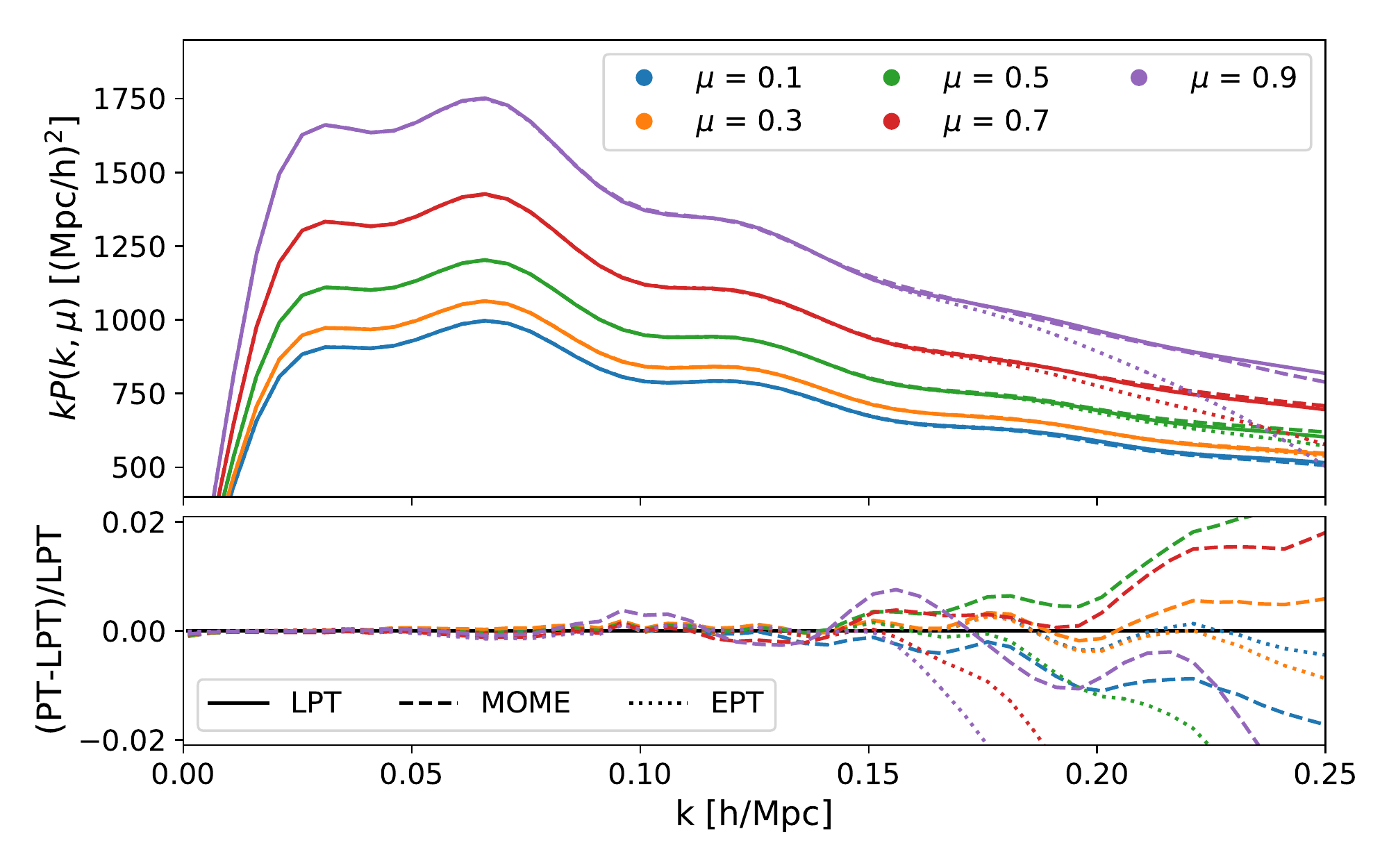}
    \caption{Power spectrum predictions of the three models (LPT, MOME, REPT) given the same set of bias parameters, but with all counterterms and stochastic parameters \edit{ adjusted in the MOME and REPT curves to match the LPT prediction with no counterterms or stochastic terms}. While much of the difference\edit{s between the three formalisms} can be soaked up by the counterterms, the fractional residuals (lower panel) at high $k$ and $\mu$ begin to show less trivial behavior, suggesting non-negligible theory error.
    }
    \label{fig:pt_wedges}
\end{figure}

Our goal in the main body of this paper was to develop the fully-resummed LPT model and compare its performance to existing models such MOME and REPT. To be as fair as possible to each of these models, we have independently fit for the bias parameters in counterterms in each. However, it should be noted that in principle the bias bases for LPT and MOME are identical, and equivalent up to a mapping to the Eulerian basis employed in REPT\footnote{See, for example, Equation 4.8 in ref.~\cite{ChenVlahWhite20}}. However, while all three models should be equivalent order-by-order under these mappings, they tend to make somewhat different predictions, especially towards small scales, due to the different IR resummation schemes involved. Namely, each of the three schemes organizes the perturbative expansion in slightly different expansion parameters. While LPT resums all the two-point long wavelength displacement and velocity contributions, in EPT these are considered perturbative and are accordingly expanded. Nevertheless, in EPT what is resummed are the contributions to the BAO feature from the relative motions of the long modes. In the MOME expansion, RSD contributions are organized following the moment expansion \cite{SelMcD11,Vla12,Vla13,VlahWhite19}, while each of the contributions is then evaluated in LPT \cite{ChenVlahWhite20}. This seemingly leaves the long velocity contributions expanded, contrary to the full LPT approach. In ref.~\cite{ChenVlahWhite20} these were left un-resummed, although a straight forward approach to add these would follow the EPT procedure, just excluding the displacement contributions which have  already been resummed. For MOME we can thus write
\begin{align}
  P^{s,IR}_{\rm 1-loop}(\bk) 
   &\approx  P^{s, {\rm nw} }_{\rm 1-loop}(\bk) + e^{-\frac{1}{2} \Sigma_s^2(\mu) k^2} \left( 1 +\tfrac{1}{2} \Sigma_s^2(\mu) k^2 \right) P^{s, {\rm w} }_{\rm lin}(\bk) 
                     +  e^{-\frac{1}{2} \Sigma_s^2(\mu) k^2} \left( P^s_{\rm loop}(\bk) - P^{s,{\rm nw}}_{\rm loop}(\bk) \right), \notag
\end{align}
where the wiggle and no-wiggle $P^{s}_{\rm 1 - loop}$ (and similarly the $P^{s}_{\rm loop}$ by dropping the linear Kaiser part) are given by MOME predictions computed in ref.~\cite{ChenVlahWhite20}. 
The difference with the EPT resummation scheme is in the definition of $\Sigma_{\rm s}$ which now contains only the velocity contributions
\begin{equation}
    \Sigma_s^2 (\mu) = f(f+2)\mu^2 \Sigma^2,
\end{equation}
where $\Sigma^2$ is the velocity dispersion due to the long wavelength modes.

Figure~\ref{fig:pt_wedges} shows the predictions for $P(k,\mu)$ of LPT, MOME and REPT when the bias parameters ($b_1$, $b_2$, $b_s$, $b_3$) are fixed to the best-fit values for the fiducial halo sample in LPT, with counterterms and stochastic contributions \edit{in the MOME and REPT cases adjusted to fit the LPT result. The three schemes differ systematically towards higher $k$ and $\mu$. Compared to its Eulerian counterpart, LPT shows significant supression of power towards high $k$, particularly along the line of sight; this suppression is absorbed by adjusting counterterms and stochastic contributions in Figure~\ref{fig:pt_wedges}, though the the theories nonetheless begin to diverge at the percent level around $k = 0.15 \kMpc$, especially towards higher $\mu$. The LPT and MOME schemes are quite similar at low $\mu$ since they differ only in the inclusion of higher-order velocities along the line of sight in the former\footnote{In fact, MOME also shows slightly \textit{more} suppression at low $\mu$ since we have followed the main text of ref.~\cite{ChenVlahWhite20} and the {\tt velocileptors} code in not including any IR cutoffs in the expanded velocity moments.}---and both are damped relative to REPT---but closer to the line of sight their oscillatory components begin to differ significantly more than between LPT and REPT since long velocity modes are not resummed in MOME.}

\edit{The above differences in the three schemes have their origin in the fact that the expansion parameters in these schemes do not match exactly. As we mentioned before, in LPT the long wavelength displacement and velocity two-point contributions are directly resummed, while in REPT only the relative effects of these are resummed and thus affect only the BAO.  The MOME scheme, on the other hand, takes a hybrid approach between the previous two. Since the predictions of the three schemes, as plotted in Figure~\ref{fig:pt_wedges}, are equal up to second (one-loop) order in the linear power spectrum when expanded order-by-order; the apparent differences therefore reflect differences at two-loop order or beyond, even though these residuals are due to the long mode contributions and are nominally under the perturbative control.  At low $k$ they manifest as contributions proportional to the wavenumber squared and can be largely absorbed by existing counterterms $\propto k^2 \mu^{2n} P(k)$ and stochastic contributions. The residual deviations at higher $k$ and $\mu$, can also provide rough estimates of the theory error of these common perturbative schemes, indicating the range of validity of current perturbative models.  We intend to return to a more in-depth comparison of these schemes in a future work.}

\begin{figure}
    \centering
    \includegraphics[width=\textwidth]{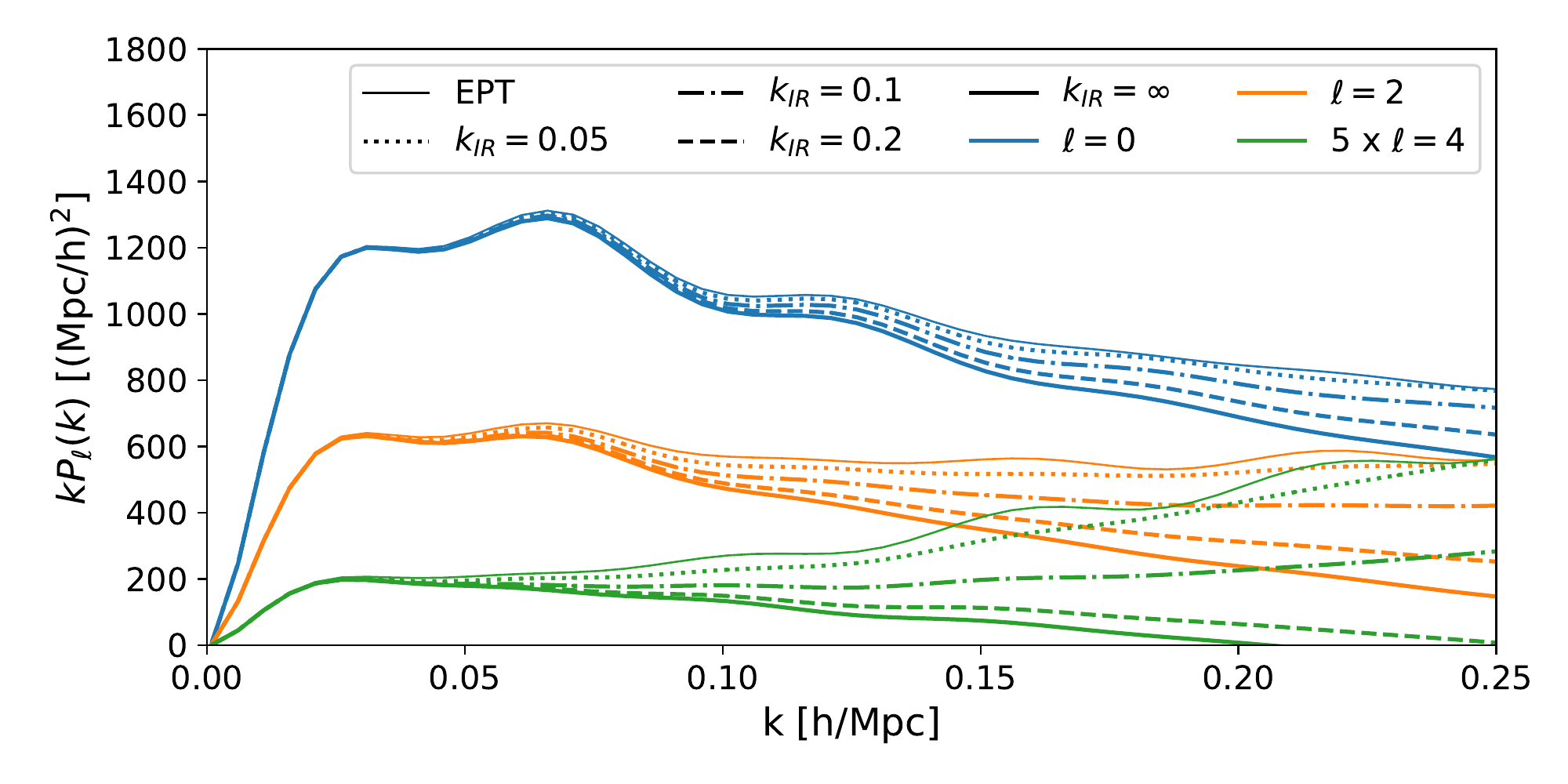}
    \caption{Power spectrum multipole predictions for a range of IR cutoffs $k_{\rm IR}$. Including more IR modes by upping $k_{\rm IR}$ damps the total power at high $k$, especially in the higher multipoles. The limit of $k_{\rm IR} \rightarrow 0$ corresponds to unresummed EPT (thin solid lines), with noticeably larger BAO oscillations at both linear and one-loop order, though even $k_{\rm IR} = 0.05 \kMpc$ dramatically reduces these.}
    \label{fig:kir_choice}
\end{figure}

In addition, as noted in the main body of the text, the choice of infrared cutoff $k_{\rm IR}$ has a significant effect on the broadband power of the anisotropic power spectrum within the LPT model itself. Figure~\ref{fig:kir_choice} shows this effect on the power spectrum multipoles, with bias parameters fixed to those obtained by fitting the fiducial sample to LPT with $k_{\rm IR} = 0.2 \kMpc$. In all three multipoles shown, decreasing the IR cutoff results in increasing power at high $k$, with higher $\ell$ increasingly sensitive to these nonlinearities. Indeed, by $k = 0.1 \kMpc$ the hexadecapoles with $k_{\rm IR}$ equal to zero is close to twice as large as that with no IR cutoff, i.e.\ $k_{\rm IR} = \infty$. In the $k_{\rm IR} \rightarrow 0$ limit the LPT prediction is equal to that in EPT without any additional IR resummation and with the bias parameters appropriately mapped. \edit{Of course, we caution that while the differences shown in Figure~\ref{fig:kir_choice} are intended to demonstrate the full spectrum of resummations possible in our scheme, in reality much of these differences can be absorbed by effective corrections as in Figure~\ref{fig:pt_wedges}.}

The increasing effect of the IR cutoff on redshift space distortions at high $\mu$ can be understood intuitively within the language of the moment expansion. In general, higher-order velocities will tend to be more sensitive to the IR cutoff. This can be seen as follows: the velocity statistics of interest for RSD can be schematically written as $M_n = \avg{X^n e^{iX}}$ where $X = \bk \cdot \Delta$. Approximating $X$ to be Gaussian with variance $\sigma^2$ we can write the even moments as 
\begin{equation}
    M_{2n} = \langle X^{2n}\rangle (1 + a_2 \sigma^2 + a_4 \sigma^4 + ...\ )
    \ \exp\left[-\sigma^2/2\right] .
\end{equation}
For $n>0$ we always have $a_2 < 0$, i.e.\ $M_{2n}$ damps faster than the exponential damping in $M_0$. This is easily understood: higher $M_n$ receive more contributions from larger values of $X$, where the complex exponential oscillates rapidly, and are suppressed by $X^n$ at small $X$ where the exponential varies slowly. 
Indeed, this effect was observed in ref.~\cite{ChenVlahWhite20}, where it was noted that the broadband of the second moment of the pairwise velocity, $\sigma_{\rm 12, ij} = \avg{(1+\delta_1)(1+\delta+2)\dDelta_i \dDelta_j}$ is very sensitive to cutoff choice, with its monopole and quadrupole respectively being better predicted by higher and lower values of $k_{\rm IR}$. Since the second moment's quadrupole is the leading $\mu^4$ contribution to the anisotropic power spectrum, one might expect that $P_4$ should in turn be very sensitive to this choice. In light of the effects of $k_{\rm IR}$ on the second moment, in this paper we have chosen the ``intermediate'' value of $0.2 \kMpc$ as our fiducial IR cutoff (at $z\approx 0.8$), though we caution that further investigation is warranted when operating at significantly higher or lower redshifts or with highly biased tracers.

\begin{figure}
    \centering
    \includegraphics[width=\textwidth]{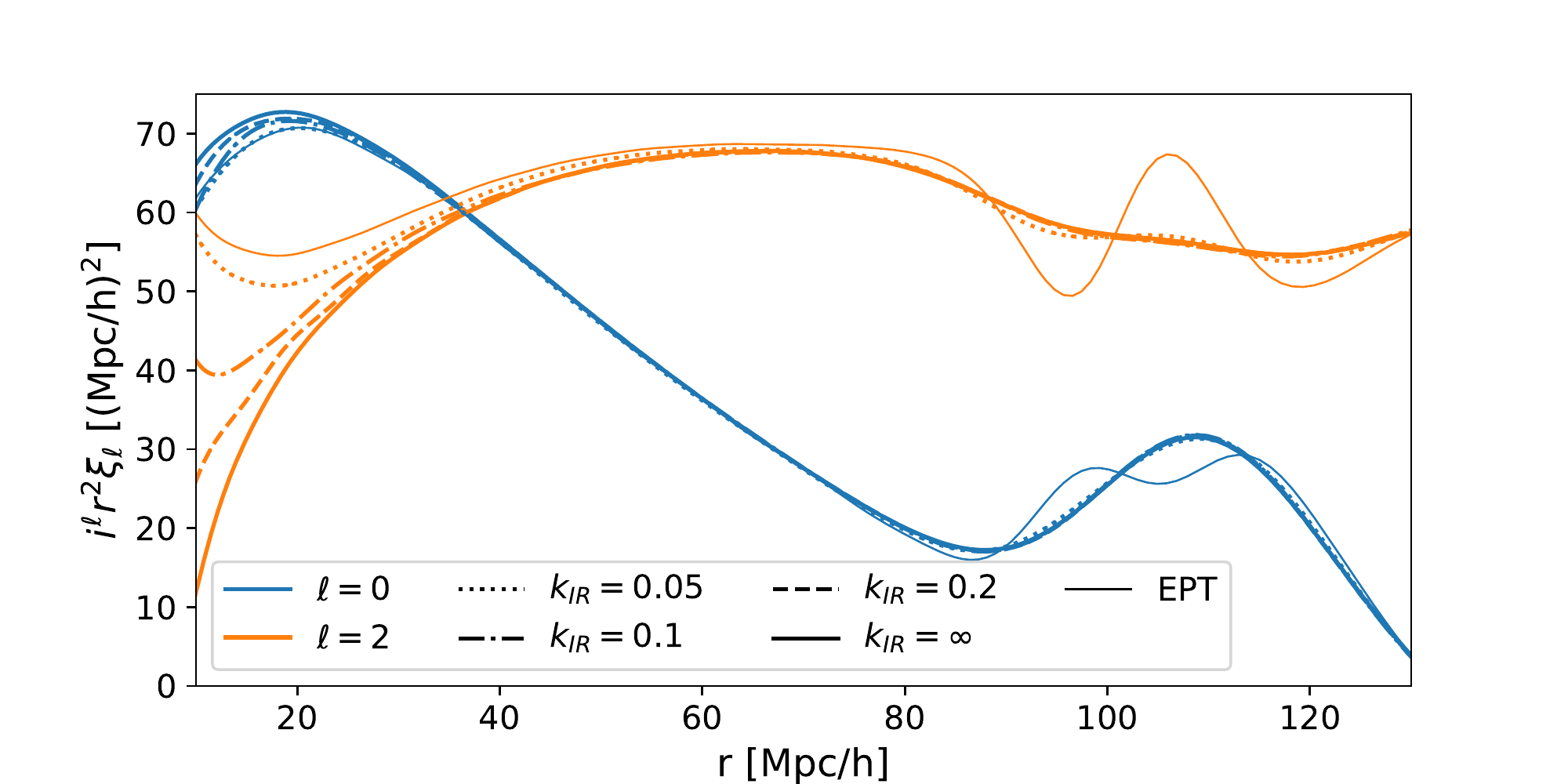}
    \caption{Correlation function multipole predictions with a range of IR cutoffs $k_{\rm IR}$ for bias parameters fixed to those obtained from the fiducial halo sample. The BAO features in both the monopole and quadrupole are rather robust, showing little change for $k_{\rm IR} > 0.05 \kMpc$, despite significant differences in the corresponding power spectrum broadbands. For reference, the unresummed EPT ($k_{\rm IR} = 0$) prediction, which clearly illustrates the non-convergence of the nonlinear configuration-space BAO feature absent IR resummation,  is also shown (thin solid lines).}
    \label{fig:xiell_kir}
\end{figure}

Finally, let us note that, in contrast to the anisotropic broadband, the corresponding BAO features in the correlation function monopole and quadrupole, shown in Figure~\ref{fig:xiell_kir}, are rather insensitive to the specific choice of $k_{\rm IR}$. Indeed, even $k_{\rm IR} = 0.05 \kMpc$, which is almost identical to EPT in its broadband, demonstrates significant damping of the BAO feature.  This figure also shows the unresummed EPT ($k_{\rm IR} = 0$) prediction, which clearly illustrates the non-convergence of the configuration-space BAO feature in one-loop EPT that necessitates the a posteriori IR resummation implemented in these models.

\section{Method I}
\label{app:mi}

In this section we extend Method I, first presented in \cite{VlahWhite19}, to include terms up to one-loop order. In contrast to Method II, described in the main body of the text, this method does not rely on boosting the wavevector $\bk$ into a more convenient frame but rather evaluates the integral in Equation~\ref{eqn:pks} directly in the observed frame.

From the form of the RSD operator
$R^{(n)}_{ij} 
= \delta_{ij} + n f \hat{n}_i \hat{n}_j,
$
where $n$ index counts a given order in PT, we can see that all integrals in Eq.\eqref{eqn:pks} can
be written in terms of scalar functions and dot products between three unit vectors ($\hq$, $\hat{k}$ and $\hat{n}$). The angular structure 
is given in terms of products 
\begin{equation}
    \hat{n} \cdot \hat{k} = \mu, \quad \hat{q} \cdot \hat{k} = \mu_{\bq}, \quad \hat{q} \cdot \hat{n} = \mu_{\bq} \mu + \sqrt{1-\mu_{\bq}^2}\,\sqrt{1-\mu^2}\,\cos\phi,
\end{equation}
where $\phi$ is the azimuthal angle in a polar coordinate system where the zenith is given by $\hat{k}$ and the plane $\phi = 0$ is spanned by $\hat{k}$ and $\hq$.
The effect of RSD operators $\mathbf{R}^{(n)}$ can then be captured by looking how it acts on the tensor basis comprised of $\hq_i, \delta_{ij},$. We have
\begin{align}
    k_i k_j \delta_{ij} &\rightarrow k_i k_j R^{(m_1)}_{in} R^{(m_2)}_{jn} = k^2 \left[1 + f(m_1+m_2+m_1 m_2f) \mu^2\right], \\
    k_i \hq_i &\rightarrow k_i R^{(m)}_{ij} \hq_j = k \mu_{\bq} \left[1 + m f\mu^2 + mf\mu^2 \gamma(\mu_{\bq},\mu) \cos\phi\right] ,
\end{align}
where  $\gamma(\mu_{\bq},\mu) = \sqrt{1-\mu_{\bq}^2}\,\sqrt{1-\mu^2}/\mu_{\bq}\mu.$ The azimuthal dependence of the exponentiated linear displacements $A^{s}_{ij}$ requires us to calculate polar-coordinate integrals of the form \cite{VlahWhite19}
\begin{equation}
    I_n(f, \mu_{\bq},\mu) = \int^{2\pi}_0 \frac{d\phi}{2\pi}\ e^{-\frac{1}{2}k^2 Y (\alpha_2 \gamma \cos\phi + \alpha_3 \gamma^2 \cos^2\phi)\mu_{\bq}^2},
\end{equation}
where $\alpha_2 = f \mu^2 (m_1+m_2 + fm_1m_2\mu^2)$ and $\alpha_3 = m_1 m_2 f^2\mu^4$. This expression can be used to compute all the 
other loop contributions (except the $W_{ijk}$ term). 
These can be calculated by taking derivatives
in either $\alpha$ or $\beta$ of the identity
\begin{equation}
I_\phi \left( \alpha, \beta, \mu_{\bq} \right) = \int_0^{2\pi} \frac{d\phi}{2\pi}~e^{ \alpha \mu_{\bq} \sqrt{1-\mu_{\bq}^2} \cos\phi
+ \beta (1 - \mu_{\bq}^2) \cos^2\phi }
=  \sum_{\ell=0}^\infty F_{\ell}(\alpha, \beta) \left(\alpha^2 \mu_{\bq}^2/\beta \right)^{\ell},~
\label{eqn:mi_phi_integral}
\end{equation}
where
\begin{align}
F_{\ell}(\alpha, \beta) &= \sum_{m=0}^\ell
\frac{\Gamma(m+\frac{1}{2})}
     {\pi^{1/2}\Gamma(m+1)\Gamma(1+2m-\ell)\Gamma(2\ell-2m+1)}
     \left( - \frac{\beta^2}{\alpha^2} \right)^m \nonumber\\
&\hspace{2.4cm} \times  M\left(\ell-2m; \ell-m+\tfrac{1}{2};  \frac{\alpha^2}{4\beta} \right)
 M\left(m+\frac{1}{2};m+1; \beta \right)
\end{align}
and $M(a,b,z)$ are hypergeometric functions of the first kind.

In order to capture the contribution of  $W_{ijk}$ a slight generalisation of the integrals above is required. 
The integral we need is of the form 
\begin{equation}
\int_0^{2\pi} \frac{d\phi}{2\pi}~
\left( \sqrt{1-\mu_{\bq}^2} \cos\phi \right) 
e^{ \alpha \mu_{\bq} \sqrt{1-\mu_{\bq}^2} \cos\phi
+ \beta (1 - \mu_{\bq}^2) \cos^2\phi }
= \frac{1}{\mu_{\bq}} \frac{d}{d \alpha} 
I_\phi \left( \alpha, \beta, \mu_{\bq} \right).
\end{equation}
However, since $F_0(\alpha, \beta)$ does not depend on 
$\alpha$ we have
\begin{equation}
\frac{1}{\mu_{\bq}} \frac{d}{d \alpha} 
I_\phi \left( \alpha, \beta, \mu_{\bq} \right)
=\mu_{\bq} \frac{1}{\beta} \frac{d}{\ d \alpha} 
\sum_{\ell=0}^\infty 
\alpha^2
F_{\ell+1}(\alpha, \beta) \left(\alpha^2 \mu_{\bq}^2/\beta \right)^{\ell}.
\end{equation}

\section{Method II}
\label{app:mii}

\subsection{General Mathematical Structure}
The workhorse integral of Method II is \cite{VlahWhite19}
\begin{equation*}
    I(A,B,C) = \int d\mu_{\bq}\ d\phi\ e^{-iC\sqrt{1-\mu_{\bq}^2}\cos\phi + iA\mu_{\bq} + B\mu_{\bq}^2} = 4\pi e^B \sum_{\ell=0}^\infty \Big( \frac{-2}{\rho} \Big)^\ell \tilde{G}^{(0)}_{0,\ell}(A,B,\rho) j_\ell(\rho),
\end{equation*}
where $\rho^2 = A^2 + C^2.$ The exact form of the kernel $\tilde{G}^{(0)}_{0,\ell}$ is given in Section~\ref{ssec:g0_defs}.

To compute the redshift-space power spectrum for biased tracers we will need the integrals derived from $A,C$ derivatives of $I$, i.e. 
\begin{align}
    I_{n,m}(A,B,C) &= i^{m-n} \int d\mu_{\bq}\ d\phi\ e^{-iC\sqrt{1-\mu_{\bq}^2}\cos\phi + iA\mu_{\bq} + B\mu_{\bq}^2} \big(\sqrt{1-\mu_{\bq}^2} \cos\phi\big)^n \mu_{\bq}^m \\
    &= 4\pi e^B \sum_{\ell=0}^\infty \Big( \frac{-2}{\rho} \Big)^\ell \tilde{G}^{(m)}_{n,\ell}(A,B,\rho) j_\ell(\rho).
    \label{eqn:Inm}
\end{align}
These satisfy the recursion relations
\begin{equation}
    \tilde{G}^{(m)}_{n,\ell} =  \frac{d\tilde{G}^{(m)}_{n,\ell}}{dA} + \frac{A}{2}\ \tilde{G}^{(m-1)}_{n,\ell-1}, \quad
    \tilde{G}^{(m)}_{n,l} = \frac{\partial \tilde{G}^{(m)}_{n,\ell}}{\partial C} + \frac{C}{2} \tilde{G}^{(m-1)}_{n,\ell-1}.
\end{equation}
For convenience we list the first few $\tilde{G}^{(m)}_{0,\ell}$:
\begin{align}
    \tilde{G}^{(1)}_{0,\ell} &=  \frac{\partial \tilde{G}^{(0)}_{0,\ell}}{\partial A} + \frac{A}{2} \tilde{G}^{(0)}_{0,\ell-1} \nonumber \\
    \tilde{G}^{(2)}_{0,\ell} &= \frac{\partial^2 \tilde{G}^{(0)}_{0,\ell}}{\partial A^2} + A \frac{\partial \tilde{G}^{(0)}_{0,\ell-1}}{\partial A} + \half \tilde{G}^{(0)}_{0,\ell-1} + \frac{A^2}{4} \tilde{G}^{(0)}_{0,\ell-2} \nonumber \\
    \tilde{G}^{(3)}_{0,\ell} &= \frac{\partial^3 \tilde{G}^{(0)}_{0,\ell}}{\partial A^3} + \frac{3A}{2} \frac{\partial^2 \tilde{G}^{(0)}_{0,\ell-1}}{\partial A^2} + \frac{3}{2} \frac{\partial \tilde{G}^{(0)}_{0,\ell-1}}{\partial A} + \frac{3A^2}{4} \frac{\partial \tilde{G}^{(0)}_{0,\ell-2}}{\partial A} + \frac{3A}{4} \tilde{G}^{(0)}_{0,\ell-2} + \frac{A^3}{8} \tilde{G}^{(0)}_{0,\ell-3}.
\end{align}
The derivatives with respect to $C$ are entirely analogous, swapping $C$ for $A$ and $m$ for $n$. In addition, we will need
\begin{equation*}
    \tilde{G}^{(1)}_{1,\ell} = \frac{\partial^2 \tilde{G}^{(0)}_{0,\ell}}{\partial A \partial C} + \frac{C}{2} \frac{\partial \tilde{G}^{(0)}_{0,\ell-1}}{\partial A} + \frac{A}{2} \frac{\partial \tilde{G}^{(0)}_{0,\ell-1}}{\partial C} + \frac{AC}{4} \tilde{G}^{(0)}_{0,\ell-2}
\end{equation*}
\begin{align*}
    \tilde{G}^{(2)}_{1,\ell} &= \frac{\partial^3 \tilde{G}^{(0)}_{0,\ell}}{\partial A^2 \partial C} + \frac{C}{2} \frac{\partial^2 \tilde{G}^{(0)}_{0,\ell-1}}{\partial A^2} + A \frac{\partial^2 \tilde{G}^{(0)}_{0,\ell-1}}{\partial A \partial C} + \half \frac{\partial \tilde{G}^{(0)}_{0,\ell-1}}{\partial C} \\
    &+ \frac{AC}{2} \frac{\partial \tilde{G}^{(0)}_{0,\ell-2}}{\partial A} + \frac{A^2}{4} \frac{\partial \tilde{G}^{(0)}_{0,\ell-2}}{\partial C} + \frac{C}{4} \tilde{G}^{(0)}_{0,\ell-2} + \frac{A^2C}{8} \tilde{G}^{(0)}_{0,\ell-3}
\end{align*}

\subsection{Example: One-Loop Matter Power Spectrum in Redshift Space}

Let us consider the one-loop matter power spectrum as an example for the kinds of angular terms that can arise. For simplicity, we focus on what happens to the \edit{un-exponentiated} one-loop contribution $A^{(22)}_{ij}$. In this case we have that the relevant quantity in redshift space is
\begin{align}
    k_i k_j (\vb{R}^{(1)} + f\ \hn \otimes \hn)_{in} (\vb{R}^{(1)} + f\ \hn \otimes \hn)_{jm} A^{(22)}_{nm} = \big(K_i K_j + 2 f k_{\parallel,i} K_j + f^2 k_{\parallel,i} k_{\parallel,j} \big) A^{(22)}_{ij}.
    \label{eqn:a22}
\end{align}
The first piece ($K_i K_j$) is identical in angular structure to those in the Zeldovich case, so we restrict our attention to the other two.

Let's begin with the term proportional to $f$ in Equation~\ref{eqn:a22}. We have (dropping the $(22)$ for brevity)
\begin{align*}
    K_i k_{\parallel,j} A_{ij} &= (K \cdot k_\parallel) X + (\hq \cdot K) (\hq \cdot k_\parallel) Y \\
    &= k^2\mu^2 (1+f)\ X + (K \mu_{\bq})\ k\mu (A(\mu) \mu_{\bq} + B(\mu)\sqrt{1-\mu_{\bq}^2} \cos\phi)\ Y.
\end{align*}
The piece proportional to $X$ poses no problem since it has no angular dependence. The term proportional to $Y$ has a piece proportional to $\mu_{\bq}^2$
, which can be computed via two derivatives of Equation~\ref{eqn:int_mii} w.r.t. $A$, and another with $\phi$ dependence calculable via a $C$ derivative; both are of the form \ref{eqn:Inm}.
The term proportional to $f^2$ is similar and involves
\begin{equation*}
    k_{\parallel,i} k_{\parallel,j} A_{ij} = k^2 \mu^2 \left[ X + (\hq \cdot \hn)^2 Y \right].
\end{equation*}
This piece proportional to $Y$ then involves up to two $C$ derivatives.

\subsection{General Angular Structure of Bias Contributions}
\label{sec:subAng}
Let us now list all the possible angular dependencies at one-loop order, organized in powers $\mu_{\bq}^a (\hn \cdot \hq)^b$. This format is convenient because each such power can be readily integrated in $\phi$ and $\mu_{\bq}$ to give
\begin{align*}
    \mu_{\bq}^a (\hn \cdot \hq)^b = \sum_{n=0}^b \binom{b}{n}  &A^n(\mu) B^{b-n}(\mu) \mu_{\bq}^{a+n} \big(\sqrt{1-\mu_{\bq}^2} \cos\phi \big)^{b-n} \rightarrow  \nonumber \\
    & 4\pi \sum_{n=0}^b \binom{b}{n}  A^n(\mu) B^{b-n}(\mu) \Big(\frac{-2}{kq}\Big)^\ell \tilde{G}_{b-n,\ell}^{a+n}\ j_\ell(kq)
\end{align*}

The simplest case involves correlators with one order $n$ displacement, of the form $U^{(n)}_i = U(q) \hq_i$:
\begin{align}
    k_i U^{s,(n)}_i =[K \mu_{\bq} + f(n-1)k \mu (\hat{q}\cdot\hat{n})]\   U(q)
\end{align}
Then we have terms involving two displacements with order $n, m$, which we can write as $A^{(n,m)}_{ij} = X \delta_{ij} + Y \hq_i \hq_j$:
\begin{align}
    k_i k_j A^{s,(n,m)}_{ij} =&  K^2 [X(q)+ Y(q) \mu_{\bq}^2 ] + (n+m-2)f k\mu [X(q)(\hat{K}\cdot\hat{n})+Y(q)\mu_{\bq}(\hat{q}\cdot\hat{n})] \nonumber \\
    & + (n-1)(m-1)f^2 k^2 \mu^2 [X(q)+Y(q) (\hat{q}\cdot\hat{n})^2].
\end{align}
Finally, at one-loop order there is one term involving three displacements involving their (112) bispectrum $W^{(112)}_{ijk} = V_1 (\hq_i \delta_{jk} + \hq_j \delta_{ik}) + V_3 \hq_k \delta_{ij} + T \hq_i \hq_j \hq_k$:
\begin{align}
    k_i k_j k_k W^{(112)}_{ijk} &= 2 \big(K^3 + f K (k\mu)^2 (1+f) \big)\mu_{\bq}\ V_1(q) \nonumber \\
    &+ K^2 \big(K\mu_{\bq} + f(k\mu)(\hn \cdot \hq) \big)\ V_3(q) \nonumber \\
    &+ K^2 \big(K\mu_{\bq}^3 + f(k\mu)\mu_{\bq}^2 (\hn \cdot \hq) \big)\ T(q).
\end{align}
The full list of non-Zeldovich angular dependences required for the one-loop power spectrum is given in Table~\ref{tab:bias_nm}.

\begin{table}
\begin{center}
\begin{tabular}{c | c }
bias & $(n,m, l)$: correlator \\ \hline
$1$ & $(1,3)$: $A^{(13)}_{ij}$, $(2,2)$: $A^{(22)}_{ij}$, $(1,1,2)$: $W^{(112)}_{ijk}$  \\
$b_1$ & $(3)$: $U^{(3)}_i$, $(1,2)$: $A^{10}_{ij}$ \\
$b_1^2$ & $(2)$: $U^{11}_i$ \\
$b_2$ & $(2)$: $U^{20}_i$ \\
$b_s$ & $(2)$: $V^{10}_i$
\end{tabular}
\caption{Contributions to the one-loop power spectrum and the perturbative order of the displacements they contain.}
\label{tab:bias_nm}
\end{center}
\end{table}

\subsection{$\tilde{G}^{(0)}_{0,m}$ and Its Derivatives}
\label{ssec:g0_defs}

The basic kernel for Method II is the function 
\begin{equation}
    \tilde{G}^{(0)}_m(A,B,\rho) = \sum_{n=m}^{\infty} f_{nm} \Big(\frac{BA^2}{\rho^2} \Big)^n  {}_2F_1\Big(\frac{1}{2}-n,-n; \frac{1}{2}-m-n; \frac{\rho^2}{A^2}\Big) ,
\end{equation}
where $\rho = \sqrt{A^2 + C^2}$, $_2F_1$ is the ordinary hypergeometric function and $f_{nm}$ is
\begin{equation}
    f_{nm} = \frac{\Gamma(m+n+\frac{1}{2})}{\Gamma(m+1)\Gamma(n+\frac{1}{2})\Gamma(1-m+n)} 
    \quad .
\end{equation}

The angular dependences in Method II require us to take $A$ and $C$ derivatives of the above. The first three derivatives of $\tilde{G}^{(0)}_0$ with respect to $A$ are given by
\begin{align}
    \frac{d\tilde{G}^{(0)}_{0,m}}{dA} = \sum_{n=m}^{\infty} \Big(\frac{BA^2}{\rho^2} \Big)^n & f_{nm} \Bigg[  \Big(\frac{2n}{A}-\frac{2nA}{\rho^2} \Big) {}_2F_1\Big(\frac{1}{2}-n,-n; \frac{1}{2}-m-n; \frac{\rho^2}{A^2}\Big) \nonumber\\ 
    + & \Big(-\frac{2\rho^2}{A^3} + \frac{2}{A} \Big) \frac{(\frac{1}{2}-n)(-n)}{(\frac{1}{2}-m-n)} {}_2F_1\Big(\frac{3}{2}-n,1-n; \frac{3}{2}-m-n; \frac{\rho^2}{A^2}\Big)  \Bigg]
\end{align}
\begin{align}
    \frac{d^2\tilde{G}^{(0)}_{0,m}}{dA^2} = \sum_{n=m}^{\infty} \Big(\frac{BA^2}{\rho^2} \Big)^n & f_{nm} \Big( \frac{\rho^2-A^2}{\rho^4} \Big) \Bigg[(2m-1-4n(m+1))\  {}_2F_1\Big(\frac{1}{2}-n,-n;\frac{1}{2}-m-n;\frac{\rho^2}{A^2}\Big) \nonumber\\
    &+ (1-4n^2+m(4n-2))\ {}_2F_1\Big(\frac{3}{2}-n,-n;\frac{1}{2}-m-n;\frac{\rho^2}{A^2}\Big)\Bigg].
\end{align}
\begin{align*}
    \frac{d^3\tilde{G}^{(0)}_{0,m}}{dA^3} =\frac{C^2}{A\rho^6} \sum_{n=m}^\infty \Big(\frac{BA^2}{\rho^2} \Big)^n & f_{nm} \Bigg[\Big( (2(1-m)(1-2m)+8n(2-m)(1+m) + 8n^2(1+m))A^2 \\
    &- (1-2m+4n(1+m) )C^2 \Big)\ {}_2F_1\Big(\half-n,-n;\half-m-n;\frac{\rho^2}{A^2}\Big) \\
    &- (1-2n)\Big( 2(1-2m+2n)(1-m+n)A^2 \\
    &- (1-2m+4n(1+m))C^2\Big)\ {}_2F_1\Big(\frac{3}{2}-n,1-n;\frac{1}{2}-m-n;\frac{\rho^2}{A^2}\Big) \Bigg]
\end{align*}

The derivatives with respect to $C$ are
\begin{align}
    \frac{d\tilde{G}^{(0)}_{0,m}}{dC} = -\frac{C}{\rho^2} \sum_{n=m}^{\infty}  \Big(\frac{BA^2}{\rho^2}\Big)^n f_{nm} \Big[& {}_2F_1\Big(\half-n,-n;\half-m-n;\frac{\rho^2}{A^2}\Big) \nonumber \\
    &- (1-2n) \ {}_2F_1\Big(\frac{3}{2}-n,1-n;\frac{1}{2}-m-n;\frac{\rho^2}{A^2}\Big)\Big]
\end{align}
\begin{align}
    \frac{d^2 \tilde{G}^{(0)}_{0,m}}{d C^2} = \rho^{-4} \sum_{n=m}^{\infty}  \Big(&\frac{BA^2}{\rho^2}\Big)^n f_{nm} \Big[ \big( (1+2m-4n(1+m))A^2 + 2C^2 \big) {}_2F_1\Big(\half-n,-n;\half-m-n;\frac{\rho^2}{A^2}\Big) \nonumber \\
    &- (1-2n) \big( (1+2m-2n)A^2 + 2C^2 \big)  {}_2F_1\Big(\frac{3}{2}-n,-n;\frac{1}{2}-m-n;\frac{\rho^2}{A^2}\Big)\Big].
\end{align}
Note that we can use $dG/dA = - (C/A)\ dG/dC$ to recast the first derivative w.r.t.\ $A$ in a convenient form as well.

In addition, we need two mixed derivatives $\partial_{C}\partial_{A}^{(1,2)} G$. These are
\begin{align}
    \frac{\partial^2 \tilde{G}^{(0)}_{0,m}}{\partial C \partial A} = -\frac{C}{A \rho^4} \sum_{n=m}^{\infty}  \Big(&\frac{BA^2}{\rho^2}\Big)^n f_{nm} \Big( (2(m-2n(1+m))A^2 + C^2) {}_2F_1(\half-n,-n;\half-m-n;\frac{\rho^2}{A^2}) \nonumber \\
    &- (1-2n)(2(m-n)A^2 + C^2){}_2F_1(\frac{3}{2}-n,-n;\frac{1}{2}-m-n;\frac{\rho^2}{A^2})
    \Big)
\end{align}
\begin{align}
    \frac{\partial^3 \tilde{G}^{(0)}_{0,m}}{\partial C \partial A^2} = \frac{C}{\rho^6} \sum_{n=m}^{\infty}  \Big(&\frac{BA^2}{\rho^2}\Big)^n f_{nm} \Big( 
    \big(2(m-2m^2-4n(1-m^2)-4n^2(1+m))A^2 \nonumber \\
    &+ 3(1-2m+4n(1+m))C^2 \big)\ {}_2F_1(\half-n,-n;\half-m-n;\frac{\rho^2}{A^2}) \nonumber \\
    &- (1-2n)\big(2(1-2m+2n)(m-n)A^2 \nonumber\\
    &+ (3-6m+8n+4mn) C^2 \big)\ {}_2F_1(\frac{3}{2}-n,-n;\frac{1}{2}-m-n;\frac{\rho^2}{A^2})
    \Big)
\end{align}

\subsection{Implementation in Python}

Our implementation of Method II, \textbf{lpt\_rsd\_fftw.py}, is available as part of {\tt velocileptors}\footnote{https://github.com/sfschen/velocileptors}, a {\sc Python} package for the one-loop redshift-space power spectrum that also includes modules implementing the moment expansion, Gaussian streaming model, and resummed Eulerian perturbation theory. The LPT module includes auxiliary functions to compute multipoles, add Alcock-Paczynski parameters not equal to one and combine the various bias contributions. We also include a sample {\sc jupyter} notebook containing example usage.

Our IR-resummation procedure is inherently rather numerically involved because of the angular dependence of the resummed displacements. To speed up this calculation (in Python) we can take advantage of the fact that the derivatives $\partial^b_C \partial^a_A\tilde{G}^{(0)}_{0,\ell}$ can be written as
\begin{equation*}
    \sum_{n=\ell}^\infty \Big[ -\half K^2 Y^{\rm lin}(q) \Big]^n (kq)^{-(a+b)} c^{2n} F_n(f,\mu),
\end{equation*}
of which the only vector operations involve multiplying by $q$ and $Y(q)$.  The remaining factors are independent of $k$ and $q$ and can be tabulated for each value of $\mu$.  Further, the only special functions we need are ${}_2F_1(\half-n,-n;\half-m-n;x)$, ${}_2F_1(\frac{3}{2}-n,-n;\frac{1}{2}-m-n;x)$, and the $\Gamma$-functions in $f_{nm}$, which can all be tabulated in advance and do not have to be calculated at each wavenumber. With these simplifications, it takes about two and a half seconds to compute the mutipoles at 50 $k$ points between $0.01$ and $0.25 \kMpc$; applying a cubic spline to interpolate between these points is sufficient to achieve sub-percent accuracy for any $k$ in this range. Similarly, it takes less than a second to compute $P(k,\mu)$ over the same number of points for a fixed $\mu$.

\subsection{An alternative formulation of Method II}

In this section we provide an alternative numerical solution to the direct evaluation of the RSD integrals in LPT. For simplicity we will present only the leading order term, \ie the Zeldovich approximation, but the main result is trivially extended to one-loop, for which we will provide the necessary ingredients.
We begin by writing the Zeldovich RSD power spectrum as 
\begin{align}
\label{eq:master}
& P_{s,\rm Zel}(k,\mu) = 2\pi  \int \dd{q} q^2 e^{-\half K^2 (X(q)+Y(q))} \int_{-1}^1 d \mu_{\vb{q}} ~  e^{i \mu_{\vb{q}} A + (\mu_{\vb{q}}^2-1) B }J_0(C \sqrt{1-\mu_{\vb{q}}}) 
\end{align}
with $A \equiv k q c$, $B\equiv -K^2 Y(q)/2$ and $C\equiv kq s$. 
We then Taylor series expand in $B$ and integrate the $A$ piece by parts $n$ times, when $n$ goes to infinity eventually, to rewrite the integral over $\mu_{\vb{q}}$ as
\begin{align}
\label{eq:master1}
  \sum_{n=0}^\infty (-1)^n\int_{-1}^1 d \mu_{\vb{q}} ~  \frac{e^{i \mu_{\vb{q}} A}}{(i A/B)^n} 2^n \frac{d^n}{d \mu_{\vb{q}}^n}\left( \frac{(\mu_{\vb{q}}^2-1)^n}{2^n n!} J_0(C\sqrt{1-\mu_{\vb{q}}^2})\right) \,.
\end{align}
Then use 10.1.48 of ref.~\cite{AS1972} to expand the $J_0$,
\begin{align}
J_0(C \sqrt{1-\mu_{\vb{q}}^2}) = \sum_{\alpha=0}^\infty (4\alpha+1) \frac{(2\alpha)!}{2^{2\alpha} (\alpha!)^2}j_{2\alpha}(C)
\leg{2\alpha}{\mu_{\vb{q}}}
\end{align}
to arrive, \edit{ using the plane wave expansion of the exponential}, at
\begin{align}
     \text{\ref{eq:master1}} &=  \sum_{n,\alpha,\ell}  (-1)^n \frac{2^n (2\alpha)!}{2^{2\alpha} (\alpha!)^2} (4\alpha+1)(2\ell+1) \frac{(i)^\ell  j_\ell(A)  j_{2\alpha}(C) }{(i A/B)^n}  \\
     \label{eq:muint}
     &\times \int_{-1}^1 d \mu_{\vb{q}} ~ \leg{\ell}{\mu_{\vb{q}}} \frac{d^n}{d \mu_{\vb{q}}^n}\left( \frac{(\mu_{\vb{q}}^2-1)^n}{2^n n!}  \leg{2\alpha}{\mu_{\vb{q}}}\right)\,.
\end{align}
Now it turns out that 
\begin{align}
 \text{\ref{eq:muint}} & = \int_{-1}^1 d \mu_{\vb{q}} ~ \leg{\ell}{\mu_{\vb{q}}} \sum_{k = 0}^n \binom{n}{k} \aleg{n}{-k}{\mu_{\vb{q}}}\aleg{2\alpha}{k}{\mu_{\vb{q}}} \\
 & = 2 \sum_{k = 0}^n \binom{n}{k} \sqrt{\frac{(2\alpha+k)!(n-k)!}{(n+k)!(2\alpha-k)!}} \tj{\ell}{n}{2\alpha}{0}{-k}{k}\tj{\ell}{n}{2\alpha}{0}{0}{0} \notag
\end{align}
in terms of 3-$j$ symbols. Putting the above equations together, the angular part of the Zeldovich RSD integral can be computed analytically 
\begin{align}
     \text{\ref{eq:master1}} &=  \sum_{n,\alpha,\ell}  (-1)^n \frac{2^n (2\alpha)!}{2^{2\alpha} (\alpha!)^2} (4\alpha+1)(2\ell+1) \frac{(i)^\ell  j_\ell(A)  j_{2\alpha}(C) }{(i A/B)^n} \\
     &\times  2 \sum_k \binom{n}{k} \sqrt{\frac{(2\alpha+k)!(n-k)!}{(n+k)!(2\alpha-k)!}} \tj{\ell}{n}{2\alpha}{0}{-k}{k}\tj{\ell}{n}{2\alpha}{0}{0}{0} \\
     & \equiv \sum_{\ell,\alpha} c_{\ell,\alpha}(k,\mu,q) j_\ell(A)  j_{2\alpha}(C)\,,
     \label{eq:mu_final}
\end{align}
and we can rewrite ZA power spectrum in redshift space as
\begin{align}
\label{eq:PkZAq}
   P_{s,\rm Zel}(k,\mu) = & 2\pi  \sum_{\ell,\alpha = 0}^\infty \int \frac{\dd{q}}{q} q^3 e^{-1/2 K^2 (X(q)+Y(q))} c_{\ell,\alpha}(k,\mu,q) j_\ell(A)  j_{2\alpha}(C) \,.
\end{align}
\edit{The two remaining sums over $\alpha$ and $\ell$ run from zero to infinity, but in practice only the first 5 terms are relevant for sub-\% precision.} Upon expanding the non oscillatory part of the integrand above in complex power laws (FFTlog) \cite{Ham00,Assassi2017},
\begin{align}
q^3 e^{-1/2 K^2 (X(q)+Y(q))} c_{\ell,\alpha}(k,\mu,q) \equiv \sum_n d_{\ell,\alpha,n}(k,\mu) q^{\nu_n} \,
\end{align}
and then using the following analytic integral
\begin{align}
    &\int \frac{\dd{q}}{q}  q^{\nu_n} j_\ell(k q c)  j_{2\alpha}(k q s) \\
    =& (k c)^{-\nu_n} \frac{\pi  2^{\nu_n-3} t^{2 \alpha} \Gamma \left(\frac{1}{2} (2 \alpha+\ell+\nu_n)\right) \, _2\tilde{F}_1\left(\frac{1}{2} (2 \alpha-\ell+\nu_n-1),\frac{1}{2} (2 \alpha+\ell+\nu_n);2 \alpha+\frac{3}{2};t^2\right)}{\Gamma \left(\frac{1}{2} (-2
   \alpha+\ell-\nu_n+3)\right)} \\
   & \equiv (k c)^{-\nu_n} I(\ell,\alpha,\nu_n,t)
\end{align}
with $t\equiv s/c$, and $_2\tilde{F}_1$ a regularized Hypergeometric function, we can perform the remaining integral in Eq.~\ref{eq:PkZAq} \footnote{\edit{See for example p.\ 401 (13.4) of ref.~\cite{Watson}. The integral exists for any $\nu_n>-2$, which is the case by an appropriate choice of the FFTlog parameters.}}. The Zeldovich RSD power spectrum can then be computed as
\begin{align}
    P_{\rm ZA}(k,\mu) = 2\pi \sum_{\ell,\alpha,n} d_{\ell,\alpha,n}(k,\mu) (k c)^{-\nu_n} I(\ell,\alpha,\nu_n,t)\,.
\end{align}
For a given set of FFTlog parameters, the function $I(\ell,\alpha,\nu_n,t)$ can be tabulated in advance and the sum above performed quickly.

Loop integrals and bias terms introduce two main complications.
First, new terms proportional to $\mu_{\vb{q}}^\beta$ appear in the angular integral in Equation \ref{eq:master}. They can be easily included by rewriting them as derivatives with respect to $A$ or $B$, analogously to what is done in the real space calculation \cite{VlaWhiAvi15}. 
Second, the loop structure will replace the $J_0$ in Equation \ref{eq:master} with more complicated functions. For the one-loop calculation at hand, we have to perform the angular integral in Equation \ref{eq:master1} with the following two terms
\begin{align}
    - i \sqrt{1-\mu_{\vb{q}}^2}\ J_1(C\sqrt{1-\mu_{\vb{q}}^2}) \quad\;\text{and}\;\quad \frac{1}{2} (1-\mu_{\vb{q}}^2) \left[ J_0(C\sqrt{1-\mu_{\vb{q}}^2}) -J_2(C\sqrt{1-\mu_{\vb{q}}^2}) \right]
\end{align}
instead of the $J_0$. Those are easy to deal with by 
noticing that
\begin{align}
     - i \sqrt{1-\mu_{\vb{q}}^2}\ J_1(C\sqrt{1-\mu_{\vb{q}}^2}) = i \partial_{C} J_0(C\sqrt{1-\mu_{\vb{q}}^2}) &  \longrightarrow i \partial_{C} j_{2\alpha}(C) \\\
     & = i \frac{2 a j_{2 a}(C)}{C}-i j_{2 a+1}(C)
\end{align}
and 
\begin{align}
     \frac{1}{2} (1-\mu_{\vb{q}}^2) \left[J_0(C\sqrt{1-\mu_{\vb{q}}^2}) -J_2(C\sqrt{1-\mu_{\vb{q}}^2})\right] & = -\partial^2_{C} J_0(C\sqrt{1-\mu_{\vb{q}}^2})  \longrightarrow - \partial^2_{C} j_{2\alpha}(C) \\
     &= - \frac{\left(4 a^2-2 a-C^2\right) j_{2 a}(C)+2 C j_{2 a+1}(C)}{C^2}\,,
\end{align}
which boil down to a reshuffling of the coefficients in the sums of Equation \ref{eq:mu_final}.

\bibliographystyle{JHEP}
\bibliography{main}
\end{document}